\documentclass[12pt,preprint]{aastex}

\slugcomment{accepted for publication in ApJ}

\shorttitle{Dust Destruction in the Early Universe}
\shortauthors{Nozawa et al.}

\begin{document}

\title{Dust Destruction in the High-Velocity Shocks \\ 
Driven by Supernovae in the Early Universe}

\author{Takaya Nozawa and Takashi Kozasa}
\affil{Division of Earth and Planetary Sciences, Graduate School 
of Science, Hokkaido University, Sapporo 060-0810, Japan; 
nozawa@ep.sci.hokudai.ac.jp, kozasa@ep.sci.hokudai.ac.jp}

\and

\author{Asao Habe}
\affil{Division of Physics, Graduate School 
of Science, Hokkaido University, Sapporo 060-0810, Japan; 
habe@astro1.sci.hokudai.ac.jp}

%\altaffiltext{1}{Visiting Astronomer, Cerro Tololo Inter-American Observatory.
%CTIO is operated by AURA, Inc.\ under contract to the National Science
%Foundation.}
%\altaffiltext{2}{Society of Fellows, Harvard University.}
%\altaffiltext{3}{present address: Center for Astrophysics,
%    60 Garden Street, Cambridge, MA 02138}
%\altaffiltext{4}{Visiting Programmer, Space Telescope Science Institute}
%\altaffiltext{5}{Patron, Alonso's Bar and Grill}

\begin{abstract}

We investigate the destruction of dust grains by sputtering in the 
high-velocity interstellar shocks driven by supernovae (SNe) in the 
early universe to reveal the dependence of the time-scale of dust
destruction on the gas density $n_{{\rm H}, 0}$ in the interstellar 
medium (ISM) as well as on the progenitor mass $M_{\rm pr}$ and 
explosion energy $E_{\rm 51}$ of SN.
The sputtering yields for the combinations of dust and ion species of 
interest to us are evaluated by applying the so-called universal 
relation with a slight modification.
The dynamics of dust grains and their destruction by sputtering in shock 
are calculated by taking into account the size distribution of each dust 
species, together with the time evolution of temperature and density of
gas in spherically symmetric shocks.
The results of calculations show that the efficiency of dust destruction
depends not only on the sputtering yield but also on the initial size 
distribution of each grain species.
The efficiency of dust destruction increases with increasing 
$E_{\rm 51}$ and/or increasing $n_{{\rm H}, 0}$, but is almost
independent of $M_{\rm pr}$ as long as $E_{\rm 51}$ is the same.
The mass of gas swept up by shock is the increasing function of 
$E_{\rm 51}$ and the decreasing function of $n_{{\rm H}, 0}$.
Combining these results, we present the approximation formula for the 
time-scale of destruction for each grain species in the early universe
as a function of $E_{\rm 51}$ and $n_{{\rm H}, 0}$.
This formula is applicable for investigating the evolution of dust
grains at the early epoch of the universe with the metallicity of 
$Z \la 10^{-3}$ $Z_\odot$.
The effects of the cooling processes of gas on the destruction of dust 
are briefly discussed.                                    

\end{abstract}


\section{Introduction}

Population III (Pop III) stars formed from metal-free gas clouds where 
H$_2$ molecule is the main coolant of gas are considered to be more 
massive than 10$^2$ $M_{\odot}$ (Bromm \& Larson 2004 and references 
therein), though there are still numerous uncertainties regarding the
mass range of the first stars due to a poor understanding of the
relevant radiative feedback in the late accretion phase (see for example 
Tan \& McKee 2004).
Once the primordial gas is enriched by dust grains, dust have great 
influences on the subsequent formation and evolution history of stars 
and galaxies; the star formation rate (SFR) in the metal-poor 
star-forming clouds is enhanced via the formation of H$_2$ molecules on 
the surface of dust grain (Hirashita \& Ferrara 2002; Cazaux \& Spaans 
2004).
In addition, the cooling of gas by thermal radiation from dust 
themselves makes even the gas clouds with metallicity as low as 
$10^{-5}$ $Z_{\odot}$ fragment into low-mass gas clumps (Schneider et
al. 2002, 2003; Omukai et al. 2005).
Thus, dust grains in the early universe may cause the characteristic
mass scale of stars to shift from very high mass to low mass of $\sim 1$ 
$M_{\odot}$ observed at present time and strongly affect the evolution
of the initial mass function (IMF) at the early epoch of the universe.

Also, dust grains in the early universe play crucial roles in revealing 
the structure and evolution of the universe from the relevant 
observations because they absorb stellar light and reemit it at infrared 
(IR)--submillimeter (submm) wavelengths. 
The thermal emission from dust at high redshifts can distort the cosmic 
microwave background (CMB) radiation (Loeb \& Haiman 1997), and the 
obscuration and reddening of starlight by dust residing in the early 
interstellar and/or intergalactic space lead to the serious 
underestimate in evaluating the cosmic SFR from the observations
directed toward the higher redshifts (e.g., Hauser \& Dwek 2001).

In fact, the recent observations have confirmed the presence of large 
amounts of dust grains at high redshifts of $z \ga 5$.
The submm (Priddey et al. 2003; Robson et al. 2004) and millimeter 
(Bertoldi et al. 2003) observations reported continuum thermal emission 
of dust grains from some quasars at $z > 5$, and have suggested a large 
amount of cold dust reaching up to $10^8$--$10^9$ $M_{\odot}$ in the 
host galaxies of quasars.
Maiolino et al. (2004a) discovered the substantial extinction of stellar 
light by dust in the broad absorption line (BAL) quasars at $z > 4.9$.
They have shown that their extinction curves could be different from
those of low-redshift quasars at $z < 4$, and have suggested that the 
difference may reflect different formation and evolution mechanisms of
dust grains at $z \ga 5$.

The main sources of dust grains in the early universe at $z > 5$ are 
considered to be supernovae (SNe) evolving from the massive stars
because of their short lifetime.
Dust grains formed in the ejecta of SNe are injected into the 
interstellar medium (ISM), and are subjected to destructive processes in 
the interstellar shocks induced by the ambient SNe (e.g., Jones et al. 
1994; Dwek, Foster, \& Vancura 1996).
Thus, the size distribution and amount of dust in the early interstellar 
space are determined by the balance between their production in SNe and 
their destruction in shocks, and change with time according to the star 
formation activity (e.g., Dwek 1998).

How much amount of dust grains are destructed in shocks as well as how 
much dust grains absorb stellar light and reemit it by thermal emission 
heavily depends on their chemical composition, size and amount.
Thus, it is essential to clarify the evolution of dust in the early
universe by treating the formation and destruction processes of dust 
grains in a consistent manner, in order to elucidate the SFR and the IMF
during the early evolution of the universe from the future observations.
Nozawa et al. (2003) investigated the chemical composition, size 
distribution and amount of dust grains formed in the ejecta of 
Pop III SNe.
In this paper, as the second step to reveal the evolution of dust grains 
in the early universe, we explore dust destruction in the high-velocity 
interstellar shock driven by the SN explosion, adopting the dust model by 
Nozawa et al. (2003) for the initial dust residing in the early ISM.

We focus on dust destruction in nonradiative shock where dust grains are 
considered to be efficiently destructed by non-thermal and thermal 
sputtering (e.g., McKee 1989) because the nonradiative shock has a high 
shock velocity ($> 100$ km s$^{-1}$) and a high gas temperature 
($> 10^6$ K).
However, only a few set of experimental data on sputtering yield are 
available for the projectile-target combinations of astrophysical
interest.
Hence, we first evaluate the sputtering yields for the combinations of 
dust species produced in Pop III SNe and ion species of interest
to us, collecting a large amount of sputtering data and applying the 
universal relation for sputtering yield proposed by Bohdansky (1984)
with a slight modification so as to reproduce well the experimental
data.
Then, we investigate the dependence of the efficiency of dust
destruction for each grain species not only on the gas density 
$n_{{\rm H}, 0}$ in the ISM but also on the explosion energy $E_{\rm 51}$
and progenitor mass $M_{\rm pr}$ of SN driving the interstellar shocks
by applying the hydrodynamical models of Pop III SNe by 
Umeda \& Nomoto (2002).
Finally, we evaluate the time-scale of destruction for each grain species.

The efficiency of dust destruction by sputtering depends on the
temperature and density of gas and the velocity of dust relative to 
gas in shock as well as the sputtering yield and the dust size.
Thus, in the calculation, we solve the time evolution of temperature and 
density of gas in spherically symmetric shocks, including the cooling 
of gas by the atomic process, the inverse Compton scattering and the 
thermal emission of dust.
In coupled with the hydrodynamical calculations for gas, we carefully 
treat the erosion of dust by thermal and non-thermal sputtering caused
by motion of dust relative to gas, taking into account the size 
distribution of each dust species.

This paper is organized as follows:
We first define the efficiency of dust destruction in \S~2. 
In \S~3, we describe the properties of the ISM in the early universe 
together with the dust model adopted in the calculations, and in \S~4, we 
present the basic equations of the hydrodynamic calculations for gas and 
the method of the simulation for dust destruction in shock.
The sputtering yield of each grain species necessary for the 
calculations of dust destruction is evaluated in \S~5, and the grain
physics in shock is described in \S~6.
The results of calculations of dust destruction are given in details for
the model of interstellar shock that is driven by a SN with 
$E_{\rm 51}=1$ and $M_{\rm pr} =20$ $M_\odot$ and is propagating into
the ISM with $n_{{\rm H}, 0}=1$ cm$^{-3}$, and the approximation formula 
for the time-scale of dust destruction as a function of $E_{\rm 51}$ and 
$n_{{\rm H}, 0}$ is derived in \S~7, where the effects of the cooling of
gas on the evolution of nonradiative shock and the destruction of dust
are briefly discussed.
The summary is presented in \S~8.

\section{The definition of the efficiency of dust destruction}

Our main goal is to reveal the time-scale of dust destruction by 
interstellar shocks in the early universe.
In general, the time-scale of destruction $\tau_{{\rm SN}, j}$ for
grain species $j$ is defined as (e.g., McKee 1989)
\begin{eqnarray}
\tau^{-1}_{{\rm SN}, j} = \epsilon_j \frac{M_{\rm swept}}{M_{\rm ISM}} 
\gamma_{\rm SN},
\end{eqnarray}
\noindent
where $\epsilon_j$ is the efficiency of destruction of grain species
$j$, $M_{\rm swept}$ is the mass of gas swept up by shock until the
shock velocity $V_{\rm shock}$ decelerates below 100 km s$^{-1}$, 
$M_{\rm ISM}$ is the mass of gas and dust in the ISM, and 
$\gamma_{\rm SN}$ is the effective SN rate.
McKee (1989) has estimated $\epsilon_j$ to be nearly constant and 
considered that $M_{\rm swept}$ is proportional to the SN explosion 
energy $E_{51}$ in units of $10^{51}$ erg.
However, it is expected that both $\epsilon_j$ and $M_{\rm swept}$
depend on not only $E_{51}$ but also the progenitor mass $M_{\rm pr}$ of 
SN and the number density of gas $n_{\rm H, 0}$ in the ISM.
Hence, we aim at deriving the dependences of $\epsilon_j$ and $M_{\rm
swept}$ on $E_{51}$, $M_{\rm pr}$ and $n_{\rm H, 0}$.
The truncation time $t_{\rm tr}$ being defined as the time at which 
$V_{\rm shock}$ drops down to 100 km s$^{-1}$, the efficiency of 
destruction $\epsilon_j$ for each dust species is calculated by
\begin{eqnarray}
\epsilon_j = \frac{M_{{\rm d}, j}^{\rm dest} (t = t_{\rm tr})}
{M_{{\rm d}, j}^{\rm swept}(t = t_{\rm tr})},
\end{eqnarray}
where $M_{{\rm d} ,j}^{\rm dest}(t)$ and $M_{{\rm d},j }^{\rm swept}(t)$
are, respectively, the mass of dust destructed and the mass of dust
swept up by shock, and the mass of gas $M_{\rm swept}$ swept up by 
spherically symmetric shock is given by
\begin{eqnarray}
M_{\rm swept} \simeq \frac{4 \pi}{3} \rho_0 R_{\rm shock}^3(t = 
t_{\rm tr}),
\end{eqnarray}
with the gas density $\rho_0$ in the ISM and the travel distance of the
shock front $R_{\rm shock}(t)$.

\section{Properties of the ISM in the early universe}

\subsection{Density, temperature, and metallicity of the ISM}

We assume that the ISM in the early universe is homogeneous and
stationary ($v=0$), referring to the results of the simulations on the 
formation of the early H II region by the first stars (Kitayama et al. 
2004), which have shown that the radiative feedback from the massive 
stars causes the ambient ISM to be homogenized with the mean gas density 
of $\la$ 1 cm$^{-3}$.
In the calculations, we consider the hydrogen number density in the ISM
to be in the range of 0.03 cm$^{-3}$ $\le n_{\rm H, 0} \le 10$ cm$^{-3}$ 
to reveal the dependence of the efficiency of dust destruction on the
gas density in the ISM.
The ISM surrounding the SN explosion is considered to be ionized by the 
radiation of the massive progenitor stars and heated to by $T_0 \sim
10^4$ K (Kitayama et al. 2004; Machida et al. 2005), and then $T_0$ is 
expected to decrease below $10^4$ K where the recombination rate can be 
high even for the metallicity as low as $Z = 10^{-4}$ $Z_\odot$ 
(Wolfire et al. 1995). 
However, the evolution of gas temperature in postshock flow is almost 
independent of $T_0$ until the end of the Sedov--Taylor phase of
high-velocity shock considered in this paper, and the temperature of gas
$T_{\rm shock}$ at the shock front does not decrease below several of 
$10^4$ K at the truncation time, as is shown in the result of
calculations (see 7.1). 
Thus, we take $T_0 = 10^4$ K as the representative gas temperature in the 
ISM, independently of the density of gas.

The truncation time as well as the time evolution of gas temperature 
and density structures in postshock flow is considered to be affected
by the cooling of gas depending on the metallicity of the ISM. 
Especially, when $T_{\rm shock} \la 10^5$ K, the efficient line coolings 
of H and He ions cause the shock velocity to decelerate rapidly, and 
resultingly regulate the truncation time.
However, the gas cooling function is almost independent of the
metallicity for $Z \la 10^{-3}$ $Z_{\odot}$ corresponding to [Fe/H] 
$\le -3$ since the metal lines do not contribute to the cooling of gas 
at $T \ge 10^4$ K (Sutherland \& Dopita 1993),
and we adopt the gas cooling function for the zero metal case given in
Sutherland \& Dopita (1993) (see 4.1).
Note that
the efficiency of dust destruction, which is defined by the mass 
ratio of dust destructed to dust swept up by shock, does not depend on 
the metallicity as long as $Z \la 10^{-3}$ $Z_{\odot}$, despite the fact 
that the amount of dust in the ISM is proportional to the metallicity,
and in the calculations we assume that the metallicity of the ISM in the 
early universe is $Z = 10^{-4}$ $Z_{\odot}$
to investigate the destruction of dust in the ISM.
The elemental composition of gas in the ISM is given in Table 1 where
the C, N, and O metals are included so as to adjust the metallicity to 
$Z = 10^{-4}$ $Z_{\odot}$ and their abundances are simply evaluated by 
multiplying the metallicity and the primordial ratios tabulated in 
Sutherland \& Dopita (1993); note that these metals play no role in the 
physical processes relevant to dust destruction because of the much 
lower abundances than H and He.
The electron number density $n_e$ is fixed to 1.128 $n_{\rm H}$ given by 
Sutherland \& Dopita (1993).

\subsection{Model of dust in the early universe}

Dust grains in the ISM at the early epoch of the universe considered in 
this paper are expected to be predominantly produced in the ejecta of
SNe.
To reveal what kind of dust grains form in the early universe, Todini \& 
Ferrara (2001) calculated dust formation in primordial core-collapse 
supernovae (CCSNe), assuming the uniform elemental composition and gas 
density within the He core.
They have shown that the newly formed grains are amorphous carbon with 
size around 300 ${\rm \AA}$, and Al$_2$O$_3$, MgSiO$_3$, Mg$_2$SiO$_4$
and Fe$_3$O$_4$ whose sizes are around 10--20 ${\rm \AA}$.
By extending the model of dust formation by Todini \& Ferrara (2001) to 
pair-instability supernovae (PISNe) whose progenitors evolve from 
metal-free stars with $M_{\rm pr}$= 140--260 $M_{\odot}$ (Umeda \&
Nomoto 2002; Heger \& Woosley 2002), Schneider, Ferrara, \& Salvattera 
(2004) found that the typical sizes of dust formed in PISNe range from 
0.001 to 0.3 $\mu$m depending on grain species.
The dust models by Todini \& Ferrara (2001) can successfully reproduce 
the extinction curve observed toward a BAL quasar SDSS1048+46 at $z=6.2$ 
(Maiolino et al. 2004b) and the IR spectral energy distribution (SED) of 
blue compact dwarf galaxy SBS 0335--052 (Takeuchi et al. 2003) except
for the slight overestimate of the far IR continuum;
SBS 0335--052 is a good target in the local universe suitable for 
scrutinizing the properties of dust formed in SNe at high redshifts
because of its young age ($< 10^7$ yr) and low metallicity (1/41 
$Z_{\odot}$) (Vanzi et al. 2000).

However, as discussed by Kozasa, Hasegawa, \& Nomoto (1989), the species 
of dust formed in the ejecta largely depend on the elemental composition 
in the He core, and their sizes are affected by the time evolution of
gas density and temperature.
Thus, Nozawa et al. (2003) extensively investigated dust production in 
Pop III CCSNe and PISNe, by considering two extreme cases for the
elemental composition in the ejecta; the unmixed case with the original 
onion-like structure and the uniformly mixed case within the He core.
In the calculations of dust formation, the radial profile of gas density 
is properly treated and the time evolution of gas temperature is
calculated by solving the radiative transfer equation including the 
energy deposition of radioactive elements, which are not taken into 
account by Todini \& Ferrara (2001).
Nozawa et al. (2003) found that a variety of grain species (C, Si, Fe, 
FeS, Al$_2$O$_3$, MgSiO$_3$, Mg$_2$SiO$_4$, SiO$_2$ and MgO) condense in 
the unmixed ejecta reflecting the difference of the elemental
composition at the formation site, and that only oxide and silicate
grains (Al$_2$O$_3$, MgSiO$_3$, Mg$_2$SiO$_4$, SiO$_2$ and Fe$_3$O$_4$) 
form in the mixed ejecta with C/O $<1$.
The average sizes of these grains span the range of 0.001--1 $\mu$m 
depending on the elemental composition and gas density at the location
of formation in the ejecta.
The species of dust formed in Pop III SNe and the behaviors of 
their size distributions are almost independent of the progenitor mass, 
but the relative mass fraction of each grain species is dependent on. 

Based on the dust models by Nozawa et al. (2003), Hirashita et al. 
(2005) have shown that the extinction curve of a quasar SDSS1048+46 at 
$z=6.2$ can be better explained by the dust grains produced in the
unmixed Pop III CCSNe than those in the mixed Pop III CCSNe.
Takeuchi et al. (2005) have also shown by using the dust models for 
$M_{\rm pr} = 20$ $M_\odot$ that only dust grains formed in the unmixed 
ejecta but not in the mixed ejecta can give an excellent fit to the 
observed SED of SBS 0335--052 over near to far IR wavelengths.
These works strongly suggest that the comparisons with the observations 
relevant to dust in the early universe prefer dust grains produced in
the unmixed Pop III CCSNe rather than those in the mixed Pop III CCSNe.

Therefore, as the representative model of the initial dust grains
residing in the early ISM, we adopt the size distribution and abundance
of each grain species obtained by the calculations of dust formation in 
the unmixed ejecta of Pop III SN with $M_{\rm pr} = 20$ $M_\odot$ by 
Nozawa et al. (2003).
For comparison, we also consider the model of dust formed in the mixed 
ejecta of the same SN.
Hereafter, the models of dust formed in the unmixed and mixed ejecta are
referred to as the unmixed grain model and the mixed grain model, 
respectively.  
The initial values of the depletion factor defined by the ratio of the 
total dust mass to the total metal mass produced in the SN is 0.224 for
the unmixed grain model and 0.297 for the mixed grain model.
The bulk density $\rho_{{\rm d}, j}$, mass fraction $A_j$ and average
size $a_{{\rm ave}, j}$ of each grain species are summarized in Table 2,
and their size spectrums are given in Figure 6a for the unmixed grain
model and in Figure 7a for the mixed grain model, where the bin width is 
0.1 dex.

It is useful for the following discussion to address here the behavior
of size distribution of each dust species produced in the unmixed
ejecta. 
C, Fe and SiO$_2$ grains have a lognormal-like size distribution with
the relatively large average radius of $> 0.01$ $\mu$m, while the size 
distribution functions of FeS, Mg$_2$SiO$_4$ and MgO grains are 
approximately power-law.
The average radii of Al$_2$O$_3$ and MgSiO$_3$ grains are very small ($< 
5 \times 10^{-3}$ $\mu$m). 
Note that Si grain shows a bimodal size distribution; the grains with
the size less than 0.03 $\mu$m occupy almost 100 \% in number but only
0.2 \% in mass.
Thus, as the average radius of Si grain, we take 0.25 $\mu$m calculated 
for the grains larger than 0.03 $\mu$m including 99.8 \% in mass.
The size distribution function summed up over all grain species is well 
fitted with a broken power-law formula whose index is $-3.5$ for the
radius larger 
than 0.06 $\mu$m and $-2.5$ for the smaller one.

\section{Physics of interstellar shock}

The deceleration rate due to drag force and the erosion rate due to 
sputtering of dust grains depend on the time evolution of gas
temperature and density in postshock flow.
In this section, we present the basic equations and initial conditions
for the hydrodynamical calculation of the time evolution of shock and
the method of the simulation for dust destruction in shock.

\subsection{The basic equations for the gas dynamics}

The hydrodynamic equations for gas to calculate the time evolution of 
spherically symmetric interstellar shock driven by a SN explosion, using 
conventional physical variables, are given by
\begin{eqnarray}
\frac{\partial \rho}{dt} + \frac{1}{r^2} 
\frac{\partial}{\partial r}(r^2 \rho v) = 0
\end{eqnarray}
\begin{eqnarray}
\frac{\partial}{dt} (\rho v) + \frac{1}{r^2}
\frac{\partial}{\partial r} (r^2 \rho v^2) = 
-\frac{\partial P}{\partial r}
\end{eqnarray}
\begin{eqnarray}
\frac{\partial U}{dt} + \frac{1}{r^2} \frac{\partial}{\partial r} 
\left[r^2 (U + P) v \right] = -\Lambda(n_{\rm H},T),
\end{eqnarray}
where $P = \rho k T/\mu m_{\rm H}$ is the gas pressure with the mean 
molecular weight $\mu$ and the hydrogen mass $m_{\rm H}$, and the sum of 
kinetic energy and thermal energy per unit volume is
\begin{eqnarray}
U = \frac{1}{2} \rho v^2 + \frac{P}{\gamma-1}
\end{eqnarray}
with adiabatic index $\gamma$, and we adopt $\gamma=5/3$ in the 
calculation. 

The cooling rate of gas $\Lambda(n_{\rm H},T)$ in units of erg s$^{-1}$ 
cm$^{-3}$ is expressed by
\begin{eqnarray}
\Lambda(n_{\rm H}, T) = n_e n_{\rm H} \Lambda_{\rm gas}(T)+
\Lambda_{\rm ic}(T) + \Lambda_{\rm d}(n_{\rm H}, T),
\end{eqnarray}
where each term represents the rate of cooling by the atomic process, 
by the inverse Compton scattering, and by the thermal emission of dust. 
In the calculations, we adopt the gas cooling function $\Lambda_{\rm
gas}(T)$ for the zero metal case given in Sutherland \& Dopita (1993)
and ignore the contribution of cooling by metal ions ejected from dust 
by sputtering, because as mentioned in 3.1, the metals do not contribute 
to the cooling of gas for the metallicity of $Z = 10^{-4}$ $Z_{\odot}$
(Sutherland \& Dopita 1993).
For the inverse Compton scattering caused by collisions of hot electrons 
in shock with CMB photons, the cooling rate $\Lambda_{\rm ic}(T)$ is 
given by (Ikeuchi \& Ostriker 1986)
\begin{eqnarray}
\Lambda_{\rm ic}(T) = 5.41 \times 10^{-32} (1+z)^4 n_e T_4
\end{eqnarray}
where $T_4$ is the gas temperature in units of 10$^4$ K, and we take
$z=20$ in the calculation.
The cooling function $\Lambda_{\rm d}(n_{\rm H}, T)$ through the thermal 
emission from dust heated by collisions with gas particles in shock is 
calculated by (Dwek 1987)
\begin{eqnarray}
\Lambda_{\rm d}(n_{\rm H}, T) = \sum_j n_{{\rm d}, j} \int 
H_j(a, T, n_{\rm H}) f_j(a) da,
\end{eqnarray}
where $n_{{\rm d}, j}$ and $f_j(a)$ are, respectively, the number
density and size distribution function of dust species $j$ at a given
$t$ and $r$. 
The collisional heating rate $H_j(a, T, n_{\rm H})$ of grains species
$j$ is given in 6.3.

The time evolution of gas temperature and density in shock driven by SN 
is characterized by the explosion energy $E_{51}$ and the progenitor 
mass $M_{\rm pr}$ which specify the structures of density and 
velocity of gas in the ejecta.
Thus, as the initial condition for interstellar shocks, we adopt the 
hydrodynamical models of the ejecta of Pop III SNe with various 
$E_{51}$ and $M_{\rm pr}$ by Umeda \& Nomoto (2002) to investigate the 
dependence of the efficiency of dust destruction on $E_{51}$ and 
$M_{\rm pr}$.
Note that in the early universe with $Z = 10^{-4}$ $Z_\odot$, the
progenitors of SNe are not only Pop III stars but also massive stars of 
the subsequent generation.
However, the structures of density and velocity in the SN with a given
$E_{51}$ and $M_{\rm pr}$ do not so much depend on the initial 
metallicity for the progenitor star whose metallicity is less than
$10^{-4}$ $Z_\odot$ (Umeda, H. 2002 in private communication).
In the calculations, we consider ordinary CCSNe with the explosion
energy of $E_{51}$=1, hypernovae (HNe) whose explosion energy is more 
than ten times that of CCSNe, and PISNe whose progenitor is very massive 
($M_{\rm pr} =$ 150, 170 and 200 $M_\odot$), and hereafter CCSNe and HNe 
with progenitor mass of $M_{\rm pr} =$ 13--30 $M_\odot$ are referred to 
as Type II SNe (SNe II).
The ejecta models used in the calculations are given in Table 3, where
the labels C, H and P represent CCSNe, HNe and PISNe, respectively, and 
the numerical value denotes the mass of the progenitor in units of solar 
mass.

\subsection{The method of simulations}

Equations (4)--(6) are solved by using the flux-splitting method
with second-order accuracy in space and first-order accuracy in time
(van Albada, van Leer, \& Roberts 1982; Mair et al. 1988).
This algorithm is one of the upwind schemes for the Euler equations and 
is well suited to solve problems involving a shock.

The calculations for the time evolution of shock and dust destruction
are performed via the following two steps.
In the first step, we assume that the SN ejecta collides with the ISM 
in ten years after the explosion and calculate the time evolution of gas 
density and temperature without including dust, based on the ejecta
models of Umeda \& Nomoto (2002). 
The spatial mesh number is 500--5000 depending on the gas density in the
ISM, and the inner 20 meshes are assigned to the ejecta. 
Note that dust destruction in this first step does not contributes to 
the evaluation of the destruction efficiency because the mass of gas
swept up by shock is less than 0.1 \% of that by $t = t_{\rm tr}$. 
Also, the cooling of gas by dust has no effects on the subsequent 
structure of shock. 
In the second step, fixing the spatial mesh number to 430, we rearrange
the density, temperature and velocity of postshock gas obtained from the 
first step into the inner 30 linear meshes. 
Then, including dust, we calculate the time evolution of shock as well
as dynamics and destruction of dust.

By treating dust as a test particle, the physical processes relevant to 
dust are calculated as follows. 
Consider a stationary ($v_{\rm d}=0$) dust grain characterized by its 
composition and number density with radius between $a$ and $a + da$ at a 
given position $R_{\rm d}$ in the ISM. 
Once the grain enters into the shock front, the relative velocity to gas 
$w_{\rm d} = v - v_{\rm d}$ is assigned. 
Then, for the temperature $T$ and density $n_{\rm H}$ of gas at the
shock front, we calculate the deceleration rate $dw_{\rm d}/dt$ due to 
drag force, the erosion rate $da/dt$ by sputtering and the heating rate 
$H(a, T, n_{\rm H})$ by collisions with gas. 
By updating the relative velocity by 
$w'_{\rm d} = w_{\rm d}+(dw_{\rm d}/dt) \Delta t$ and the grain radius
by $a'=a+(da/dt)\Delta t$ for a given small time step $\Delta t$, the 
velocity and position of the dust at time $t'=t + \Delta t$ are 
calculated by $v'_d = v+ w'_d$ and 
$R'_{\rm d} = R_{\rm d} + v'_{\rm d} \Delta t$.  
The procedure described above is repeated for the gas temperature $T$ 
and density $n_{\rm H}$ at the position $R'_{\rm d}$ of the grain. 
Note that the dust grain is considered to be completely destroyed if 
the grain radius is smaller than the nominal monomer radius of
condensate given in Nozawa et al. (2003).

\section{Evaluation of sputtering yield}

Sputtering is divided into two categories reflecting the difference in 
the underlying mechanisms; physical sputtering and chemical sputtering.
Physical sputtering invokes a transfer of kinetic energy from the
incident particle to target atoms and subsequent ejection of those atoms 
which have acquired kinetic energy enough to overcome the binding forces 
exerted by the target.
Chemical sputtering invokes a chemical reaction induced by the impinging
particles which produces an unstable compound at the target surface 
(Sigmund 1981).
In the high-velocity shocks considered in this paper, dust grains are 
predominantly destroyed by physical sputtering because dust temperature  
is at most 100 K in the postshock region (Dwek 1987); chemical
sputtering is of no importance since this process is realized at the 
temperature as high as 800 K (Roth 1983).
Thus, in what follows, we simply refer to physical sputtering as 
"sputtering".
The efficiency of destruction by sputtering is characterized by the 
sputtering yield defined as the mean number of emitted atoms per 
incident particle.
The sputtering yield depends on the impact energy of a projectile and
the incident angle to the target surface as well as on the
projectile-target combination. 
However, only a few set of experimental data on sputtering yield are 
available for the combinations of dust materials and ion species of 
interest to us. 

One method evaluating the sputtering yields for the projectile-target 
combinations of astrophysical interest is to apply
the Monte-Carlo code for the 
\textit{TRANSPORT OF IONS IN MATTER} (TRIM, Ziegler, Biersack, \& 
Littmark 1985) that simulates the energy loss of a projectile ion with 
given impact energy and incident angle in a solid 
(Field et al. 1997;  May et al. 2000; Bianchi \& Ferrara 2005).
The TRIM code requires as input parameters not only the surface binding 
energy of the target, but also the displacement energy $E_{\rm d}$ and 
the lattice binding energy $E_{\rm b}$.
Particularly, uncertainties in the value of $E_{\rm d}$ are a
significant source of error in the calculated sputtering yield 
(May et al. 2000; Bianchi \& Ferrara 2005).
In fact, the sputtering yields computed by this simulation code have 
been known to significantly differ from the experimental data at low 
impact energies (Bianchi \& Ferrara 2005).
Furthermore, it is time-consuming to combine the Monte-Carlo simulation 
with the hydrodynamical calculation for dust destruction.  

Another one is to apply the universal relation which is the analytic 
formula derived by Bohdansky (1984) for the energy dependence of 
sputtering yield at normal incidence for a monatomic solid.
To investigate the destruction of interstellar dust such as C, Fe, SiC, 
SiO$_2$ and H$_2$O ice, Tielens et al. (1994) have applied the universal 
relation by treating a constant $K$ in this formula as a parameter, and 
determined it by comparing the available experimental data with the 
sputtering yield calculated by the universal relation. 
They have shown that the universal relation can be applied to not only 
monatomic solids but also compound solids by taking the appropriate 
average values for the atomic and mass numbers and the surface binding 
energy under the assumption that each of element consisting of the
target is sputtered off with the same rate.
However, the agreement of the calculated sputtering yield with the 
experimental data is not so much good for the ion-target combinations
with the mass ratio of target atom to incident ion ranging from 0.7 
and 2.

Therefore, we apply the universal relation for sputtering yield with a 
slight modification so as to reproduce well the experimental data, 
collecting and examining minutely a large amount of sputtering data, in
order to evaluate the sputtering yields for the combinations of dust and 
ion species of interest to us.
Following Tielens et al. (1994), we treat $K$ as a free parameter and 
determine it by fitting to the experimental data for the grain species
with the yield data covering a wide range of impact energy of the 
projectiles. 
For the grain species for which little or no sputtering data are 
available, we deduce the value of $K$ by fitting to the sputtering yield
calculated by means of the Monte-Carlo code for the \textit{EROSION AND 
DEPOSITION BASED ON DYNAMIC MODEL} (EDDY, Hirooka 2001).
The EDDY code requires only the surface binding energy as an input
parameter and allows us to predict the sputtering yield for specific 
impact energy and incident angle more easily than the TRIM code, though 
the energy range applicable to the calculations is limited to 0.1--10
keV.
In the following subsection, we at first introduce the universal
relation for sputtering yield, and then determine the value of $K$ for 
each grain species by fitting to the available experimental data and/or
the results of the EDDY simulation.

\subsection{Universal relation for sputtering yield}

For backward sputtering and with the neglect of inelastic energy losses,
the sputtering yield at normal incidence $Y_i^0(E)$ by the projectile
$i$ impacting with energy $E$ is given by (Bohdansky 1984)
\begin{eqnarray}
Y^0_i(E) = 4.2 \times 10^{14} \frac{S_i(E)}{U_0} 
\frac{\alpha_i(\mu_i)}{K \mu_i + 1}
\left[1- \left( \frac{E_{\rm th}}{E} \right)^{2/3} \right]
\left(1- \frac{E_{\rm th}}{E} \right)^2,
\end{eqnarray}
where $U_0$ is the surface binding energy in units of eV,
$\mu_i=M_{\rm d}/M_i$ is the ratio of the mass number of the target atom
to the incident ion, $\alpha_i$ is the energy-independent function of 
$\mu_i$, and $K$ is considered to be a free parameter to be adjusted to
reproduce the experimental data (Tielens et al. 1994).
The threshold energies $E_{\rm th}$ have been obtained by fitting the 
yield data for low-energy sputtering and are approximately given by
(Bohdansky et al. 1980; Andersen \& Bay 1981)
\begin{eqnarray}
E_{\rm th} = \frac{U_0}{g_i(1-g_i)}~~~~~~~~~~{\rm for}~~
\frac{M_i}{M_{\rm d}} \le 0.3 \nonumber \\
E_{\rm th} = 8 U_0 \left( \frac{M_i}{M_{\rm d}} \right)^{1/3}~~~
{\rm for}~~\frac{M_i}{M_{\rm d}} > 0.3,
\end{eqnarray}
where $g_i = 4 M_i M_{\rm d}/(M_i + M_{\rm d})^2$ is the maximum 
fractional energy transfer possible in a head-on elastic collision.
The function $S_i(E)$ is the nuclear stopping cross section in units of 
ergs cm$^2$ and can be expressed by the following universal relation 
(Sigmund 1981)
\begin{eqnarray}
S_i(E) = 4 \pi a_{\rm sc} Z_i Z_{\rm d} e^2 \frac{M_i}{M_i+M_{\rm d}} 
s_i(\epsilon_i),
\end{eqnarray}
where $e$ is the elementary charge, and $Z_i$ and $Z_{\rm d}$ are the 
atomic numbers of the projectile and the target, respectively.
The screening length $a_{\rm sc}$ for the interaction potential between 
the nuclei is given by
\begin{eqnarray}
a_{\rm sc} = 0.885 a_0 \left(Z_i^{2/3} + Z_{\rm d}^{2/3} \right)^{-1/2}
\end{eqnarray}
with the Bohr radius $a_0=0.529$ \AA.
The function $s_i(\epsilon_i)$ depends on the detailed form adopted for 
the screened Coulomb interaction and can be approximated by
(Matsunami et al. 1980)
\begin{eqnarray}
s_i(\epsilon_i) = \frac{3.441 \sqrt{\mathstrut \epsilon_i} 
\ln(\epsilon_i+2.718)}{1+6.35 \sqrt{\mathstrut \epsilon_i} + 
\epsilon_i \left(-1.708 + 6.882 \sqrt{\mathstrut \epsilon_i} \right)},
\end{eqnarray}
where the reduced energy $\epsilon_i$ is given by
\begin{eqnarray}
\epsilon_i = \frac{M_{\rm d}}{M_i+M_{\rm d}} \frac{a_{\rm sc}}{Z_i Z_{\rm d} 
e^2} E.
\end{eqnarray}
The value of $\alpha_i(\mu_i)$ in Equation (11) depends on how the 
distribution of energy deposited in the target is approximated.  
Sigmund (1969) suggested the three types of approximations
according to the adopted distribution function (Gaussian, corrected
Gaussian, and non-Gaussian).
The values of $\alpha_i(\mu_i)$ obtained by these approximations are 
very similar (Sigmund 1969) and is approximately expressed for 
$\mu_i \le 10$ as follows (e.g., Bohdansky 1984); 
\begin{eqnarray}
\alpha_i &=& 0.2~~~~~~~~{\rm for}~~\mu_i \le 0.5
\nonumber \\
\alpha_i &=& 0.3 \mu_i^{2/3}~~~{\rm for}~~0.5 < \mu_i \le 10.
\end{eqnarray}
However, Equation (17) for $\alpha_i$ slightly overestimates the 
sputtering yields for the projectile-target combinations with the mass 
ratio of $0.7 \la \mu_i \la 2$.
Hence, from a comparison with a large amount of sputtering data, we
apply the following formula for $\alpha_i$; 
\begin{eqnarray}
\alpha_i &=& 0.2~~~~~~~~~~~~~~~~~~~~~~~~~~~~~~~~{\rm for}~~\mu_i \le 
0.5 \nonumber \\
\alpha_i &=& 0.1 \mu_i^{-1} + 0.25 \left( 
\mu_i-0.5 \right)^2~~~
{\rm for}~~0.5 < \mu_i \le 1 \nonumber \\
\alpha_i &=& 0.3 \left( \mu_i - 0.6 \right)^{2/3}
~~~~~~~~~~~~~~~{\rm for}~~1 < \mu_i.~~~~~~~
\end{eqnarray}
As we shall show below, this modified function $\alpha_i$ can give 
an excellent agreement with the sputtering data for any values of
$\mu_i$ considered here ($0.3 \le \mu_i \le 56$).

By being $K$ as a parameter, the surface binding energy $U_0$ must be
specified to evaluate the normal-incidence sputtering yield $Y^0_i(E)$
from Equation (11).
We assume $U_0$ to be equal to the sublimation energy and evaluate it 
from the JANAF Thermochemical Tables (Chase et al. 1985) except for C
and Al$_2$O$_3$ whose values of $U_0$ are discussed later.
The values of $U_0$ for each grain species are summarized in Table 4
along with the values of $Z_{\rm d}$ and $M_{\rm d}$.

\subsection{Sputtering yield of each grain species}

In order to evaluate the sputtering yield for the projectile-target 
combinations of interest to us, we determine the value of $K$ by fitting 
to the experimental data for C, Si, Fe, SiO$_2$ and Al$_2$O$_3$ grains for 
which a sufficient amount of sputtering data are available.
For MgO, FeS and Fe$_3$O$_4$ grains for which little or no yield data
are available, we evaluate the value of $K$ by fitting to the results of
the EDDY simulation.
The experimental and simulated sputtering data at normal incidence, for
which the references are summarized in Table 4, are given in Figure 1 along 
with the best-fitting theoretical yield calculated by the universal relation
(solid lines). 
In Figures 1a--1e, the experimental data are shown for the following 
ion species; H$^+$ (open circles), D$^+$ (asterisks), He$^+$ (open 
squares), Ne$^+$ (crosses) and Ar$^+$ (open triangles).
Even in the case that the experimental data can be sufficiently 
available, for comparison, we also plot the results of the EDDY 
calculations for the projectiles H$^+$ (filled circles), He$^+$ (filled 
squares) and Ar$^+$ (filled triangles).

As for C, the sublimation energy estimated to be $U_0=7.43$ eV from the 
thermodynamical data cannot reproduce the experimental data. 
Tielens et al. (1994) have shown that the ion fluences efficiently 
amorphize the surface layer of carbon material and reduce the surface 
binding energy.
Thus, we take $U_0=4.0$ eV following Tielens et al. (1994), and this
value gives a good fit to the measured yields with $K=0.61$ (Fig. 1a)
being different from $K=0.65$ by Tielens et al. (1994); the difference 
stems from the adopted formula for $\alpha_i$.
The sputtering yield calculated by the EDDY code is enhanced for H$^+$ by a
couple of factor compared to the experimental data, but gives a good 
agreement for He$^+$.

The sputtering data for Si and Fe targets are sufficiently available.
The modified universal relation employing Equation (18) for $\alpha_i$ 
shows the excellent agreement with the 
data points for all projectiles considered here (Figs. 1b and 1c), 
by adopting $K=0.43$ and $K=0.23$, respectively. 
The EDDY calculations for Si and Fe targets also present a good
agreement with their measured data.
The dotted lines indicate the sputtering yields by Ar$^+$ projectile 
calculated by using the values of $K$ derived above but adopting 
$\alpha_i$ given by Equation (17).
It can be seen from these figures that Equation (17) 
overestimates the sputtering yields for the projectile-target 
combinations with the mass ratio of $0.7 \la \mu_i \la 2$ as mentioned 
in 5.1. 
It should be emphasized that the previous
formula  
cannot reproduce the 
experimental data better than the modified universal relation whatever  
values of $K$ are selected.

A very good fit of the universal relation to the experimental data is 
obtained for SiO$_2$ target by taking $K=0.1$ (Fig. 1d).
This justifies the application of the universal relation to compound
materials as shown by Tielens et al. (1994).
The results by the EDDY simulations are consistent with the experimental 
data, but a comparison with the theoretical curves of sputtering yield 
shows a little enhancement for H$^+$ and Ar$^+$ projectile at $E \le 200$ 
eV.
For MgSiO$_3$ and Mg$_2$SiO$_4$ materials for which the sputtering data 
cannot be found in the literatures, as is the same as Tielens et al. 
(1994), we assume $K$ to be 0.1, considering SiO$_2$ as a
representative of silicates.
In fact, this could be reasonable since $K \simeq 0.1$ for the 
compounds including oxygen atoms (see Table 4).

Given the surface binding energy of $U_0=6.37$ eV evaluated from the 
thermodynamical data, the yield data of Al$_2$O$_3$ cannot be reproduced 
by the universal relation for any values of $K$.
The disagreement of the theoretical prediction with the yield data is
expected to be due to an unsuitable estimate of $U_0$, as is the same
as C.
Indeed, Roth, Bohdansky, \& Ottenberger (1979) have suggested $U_0=8.5$ 
eV for Al$_2$O$_3$ from the behavior of sputtering yield    
at impact energies below the maximum yield by He$^+$.
Adopting this surface binding energy, we can get a better fit with 
$K=0.08$ (Fig. 1e), although the agreement of the universal relation
with the yield data is the worst among grain materials considered in
this paper.

To our knowledge, little or no experimental data on sputtering yield 
exist for MgO, FeS and Fe$_3$O$_4$.
Thus, we employ the results by the EDDY calculations to extract the 
values of $K$ for these materials.
In the figures from 1f to 1h (1f for MgO, 1g for FeS, and 1h for 
Fe$_3$O$_4$), we show the EDDY results and the best-fitting theoretical 
curves.  
The agreements are not always good for H$^+$ and Ar$^+$ at low energies, 
which is also true for SiO$_2$, but the universal relation well
reproduces the yield data of SiO$_2$. 
Thus, we consider that the fits are reasonable and satisfactory for 
these materials. 
The values of $K$ determined are 0.06, 0.18 and 0.15 for MgO, FeS and 
Fe$_3$O$_4$, respectively.

Table 4 summarizes the value of $K$ for each grain species evaluated
by fitting to the available experimental data and/or the results of the 
EDDY simulations.
We confirm that the derived value of $K$ can well reproduce the 
experimental data of compounds as well as single-element materials, 
by adopting appropriate average values for the atomic and mass numbers
as proposed by Tielens et al. (1994). 
By slightly modifying the function $\alpha_i(\mu_i)$ in the universal 
relation, we realized that the calculated sputtering yields well agree  
with the available experimental data than before.
The modified universal relation derived here could be applicable to 
evaluate the sputtering yield for any combinations of targets and 
projectiles of astrophysical interest as long as the experimental data
are available.

\section{Physics of dust grains in shock}

The collisional interactions between dust and a hot gas in shock 
efficiently transfers charge, momentum and energy between the two 
phases.
As a result, dust grains acquire the electric charge, undergo the 
resistance to their motion through the gas, are eroded by sputtering,
and are heated by collisions with gas.
Also, heated dust grains cool the gas through their thermal emission.

Electric charge on a grain plays an important role in grain physics in
hot plasmas through the Coulomb interaction and the Lorentz force in the 
presence of magnetic field. 
Here, we concentrate on the destruction of dust grains in  fast 
($\ge 100$ km s$^{-1}$) and hot ($\ge 10^5$ K) nonradiative shocks.
For $T \ga 10^6$ K, the effect of the electric charge of dust grain can
be neglected because the dimensionless potential parameter $\phi$ 
defined by the ratio of the electric potential acquired by dust to 
the kinetic energy of gas is approximated by $\phi \sim 10^5/T$ and is 
significantly smaller than unity (Draine \& Salpeter 1979; McKee et al. 
1987). 
Also, the erosion rate of refractory grains by thermal sputtering 
quickly decreases for $T < 10^6$ K (Draine \& Salpeter 1979). 
Correspondingly, the erosion rate of dust grans by non-thermal
sputtering decreases quickly for the relative velocity of dust to gas 
$w_{\rm d} < 200$ km s$^{-1}$. 
Therefore, it can be expected that the electric charge does not 
significantly affect the destruction of dust grains in nonradiative 
shocks considered in this paper. 
Also, it is assumed that the early ISM is not pervaded by a magnetic
field because the magnetic field in the early universe is considered to
be very weak (Gnedine, Ferrara, \& Zweibel 2000). 
Thus, we formulate the basic equations describing the motion of dust 
grains and their erosion by sputtering in postshock flow, neglecting the 
effect of charge on dust grain.

\subsection{Dynamics of dust grain}

Once dust grains being at rest in the preshock ISM encounter the 
interstellar shock, they move ballistically behind the shock front with 
the initial velocity relative to gas $w_{\rm d} \simeq (3/4) V_{\rm
shock}$ where $V_{\rm shock}$ is the shock velocity, and are eroded by
non-thermal sputtering.
A grain streaming through the ionized gas is decelerated by a drag force 
due to the direct collisions with gas particles, and thus the velocity
of dust relative to gas decreases.
The drag force acting on a grain results from the momentum transfer from 
gas to dust.
Under the assumption that a gas particle reflects specularly from grain 
surface and that the kinetic temperature of gas is the same for all gas
species, the deceleration rate of a spherical grain of radius $a$ with 
the velocity relative to gas $w_{\rm d}$ is given by (e.g., Draine \&
Salpeter 1979)
\begin{eqnarray}
\frac{d w_{\rm d}}{dt}=-\frac{3 n_{\rm H}k T}{2 a \rho_{\rm d}} 
\sum_i A_i G_i(s_i),
\end{eqnarray}
where $n_{\rm H}$ is the number density of hydrogen atom, 
$k$ is the Boltzmann constant, $T$ is the gas temperature, 
$\rho_{\rm d}$ is the bulk density of grain and 
$s_i^2 = m_i w_{\rm d}^2/2kT$ with the mass $m_i$ of gas species $i$. 
The summation is taken over all gas species, and $A_i$ is the number 
abundance of gas species $i$ relative to that of hydrogen atom.
For the function $G_i(s_i)$ whose exact formula is given by Baines, 
Williams, \& Asebiomo (1965), the following analytical approximation 
has been proposed by Draine \& Salpeter (1979)
\begin{eqnarray}
G_i(s_i) \approx \frac{8 s_i}{3 \sqrt{\pi}} \left(1 + \frac{9 \pi}{64} 
s_i^2 \right)^{\frac{1}{2}}
\end{eqnarray}
with the accuracy within 1 \% for $0 < s_i < \infty$.
In the calculations, we adopt this formula.
It can be seen from Equation (19), grains with smaller size and/or 
higher relative velocity can be effectively decelerated by drag force
and come to comove with gas in shock.

\subsection{Grain destruction by sputtering}

The destruction of dust grains in high-velocity and high-temperature
nonradiative shock is dominated by sputtering.
Actually, we can neglect the thermal evaporation of dust, because the 
temperature of dust cannot be raised as high as their sublimation 
temperature ($\sim$ 1000--2000 K) for the preshock gas density
considered in this paper (e.g., Dwek 1987).
Furthermore, we neglect the shattering process and partial vaporization
caused by grain-grain collisions, since the number density of dust 
particles is very small in postshock flow and collisions between dust 
grains are extremely rare without a magnetic field.

The erosion rate of grain by sputtering is calculated by taking the 
angle-averaged sputtering yield $\langle Y_i(E_i) \rangle_\theta = 2 
Y_i^0(E_i)$ (Draine \& Salpeter 1979) 
and is given by (e.g., Dwek et al. 1996)
\begin{eqnarray}
\frac{da}{dt} = - \frac{m_{\rm sp}}{2 \rho_{\rm d}} n_{\rm H}
\sum_i A_i \left( \frac{8 k T}{\pi m_i}\right)^{\frac{1}{2}} 
\frac{e^{-s_i^2}}{2s_i}
%\nonumber \\ \times 
\int \sqrt{\epsilon_i} e^{-\epsilon_i} 
\sinh(2 s_i \sqrt{\epsilon_i}) Y_i^0(\epsilon_i) d\epsilon_i
\end{eqnarray}
where $\epsilon_i=E_i/kT$ and $m_{\rm sp}$ is the average mass of the 
sputtered atoms.
Equation (21) can be reduced to the following equations for the two 
extreme cases of $s_i$ (Dwek \& Arendt 1992; Tielens et al. 1994);
one is for $s_i \ll 1$ where a stationary grain suffers the thermal 
sputtering caused by collisions originating from thermal motion of gas, 
and the erosion rate is written as
\begin{eqnarray}
\frac{1}{n_{\rm H}} \frac{da}{dt} =-  \frac{m_{\rm sp}}{2 \rho_{\rm d}}
\sum_i A_i \left( \frac{8 k T}{\pi m_i} \right)^{\frac{1}{2}} \int 
\epsilon_i e^{-\epsilon_i} Y_i^0(\epsilon_i) d\epsilon_i.
\end{eqnarray}
The other is in the limit of $s_i \rightarrow \infty$ where the 
non-thermal sputtering erodes a hypersonic grain with the rate given by 
\begin{eqnarray}
\frac{1}{n_{\rm H}} \frac{da}{dt} =-  \frac{m_{\rm sp}}{2 \rho_{\rm d}} 
w_{\rm d} \sum_i A_i Y_i^0(E=0.5 m_i w_{\rm d}^2).
\end{eqnarray}

Figure 2 shows the erosion rate of each dust species by sputtering in 
the gas with the primordial elemental composition of metallicity $Z =
10^{-4}$ $Z_\odot$ given in Table 1; 
Figure 2a for thermal sputtering versus the gas temperature $T$ from
Equation (22) and Figure 2b for non-thermal sputtering versus the 
velocity of dust relative to gas $w_{\rm d}$ from Equation (23).
The erosion rate by thermal (non-thermal) sputtering steeply increases 
from $T \sim 10^5$ K ($w_{\rm d} \sim 30$ km s$^{-1}$), reaches a peak
at $T=$ 4--20 $\times 10^7$ K ($w_{\rm d} = 500$--1300 km s$^{-1}$), and 
then slowly decreases with increasing $T$ ($w_{\rm d}$).
This behavior of the erosion rate as a function of $T$ and $w_{\rm d}$ 
reflects the dependence of sputtering yield on impact energy (see 5.2).
Among dust species considered in this paper, C grain has the lowest
erosion rate at $T \ge 2 \times 10^6$ K ($w_{\rm d} \ge 200$ km
s$^{-1}$), which is about one order of magnitude lower than that of FeS 
grain with the highest rate at $T \ge 10^7$ K ($w_{\rm d} \ge 400$ km 
s$^{-1}$).
For thermal sputtering, the erosion rate $da/dt$ for the other dust 
species is $\sim 1.2 \times 10^{-6} n_{\rm H}$ $\mu$m yr$^{-1}$ cm$^3$
within a factor of three at $T \ge 2 \times 10^6$ K.
For non-thermal sputtering at $w_{\rm d} \le 200$ km s$^{-1}$, the 
destruction by He$^+$ is dominant, and the erosion rate of Al$_2$O$_3$
is the lowest.

\subsection{Grain heating}

The collisions with gas can also heat dust grains in postshock flow.
The collisional heating rate $H(a, T, n_{\rm H})$ of dust grain
with radius $a$ is presented by Dwek \& Arendt (1992) as 
\begin{eqnarray}
H(a, T, n_{\rm H}) = n_{\rm H} \pi a^2 k T \sum_i A_i 
\left( \frac{8 k T}{\pi m_i}\right)^{\frac{1}{2}} \frac{e^{-s_i^2}}{2s_i}
\int \epsilon_i^{\frac{3}{2}} e^{-\epsilon_i} 
\sinh(2 s_i \sqrt{\epsilon_i}) \eta_i(a, \epsilon_i) d\epsilon_i
\end{eqnarray}
where $\eta_i(a, \epsilon_i)$ is the fraction of kinetic energy 
of gas species $i$ deposited into the dust grain.
The values of $\eta_i(a, \epsilon_i)$ are calculated by comparing
an effective grain thickness $a_{\rm eff}$ ($4 a/3$ for a spherical
grain) with the stopping range of the incident particle $l_{\rm s}$.
When $a_{\rm eff} \ge l_{\rm s}$, the incident particle deposits almost 
all of the energy into dust grains, otherwise, the difference between 
$a_{\rm eff}$ and $l_{\rm s}$ is exploited to determine the fraction of 
deposited energy.
The approximations for the stopping range of ion species considered here 
are given in Dwek (1987).
For the electron stopping range, we adopt the approximation derived by 
Tabata, Ito, \& Okabe (1972) and Iskef, Cunningham, \& Watt (1983).
This electron stopping range reduces the heating rate by $\sim$ 30 \%
at $10^6$ K $\la T \la 10^9$ K compared with that given by Dwek (1987).
The values and detailed derivation of $\eta_i(a, E)$ will be presented
elsewhere (Nozawa et al. in preparation).

\section{Results of calculation of dust destruction}

In this section, we show the results of numerical simulations for dust 
destruction in nonradiative shock, focusing on the results calculated
for the ISM-dust specified by the unmixed grain model described in \S~3.
In 7.1, we demonstrate the time evolution of temperature and density of
gas and the destruction of dust grains in the interstellar shock that is 
driven by the SN model of C20 with $E_{\rm 51}=1$ and $M_{\rm pr}=20$ 
$M_\odot$ and is propagating into the ISM with $n_{{\rm H}, 0}=1$ 
cm$^{-3}$.
In 7.2, we investigate the dependences of the destruction efficiency and 
the mass of gas swept up by shock on $E_{51}$, $M_{\rm pr}$ and 
$n_{\rm H, 0}$, and derive the approximation formula for the time-scale
of destruction for each dust species in the early universe.
The effects of the cooling processes of gas on the evolution of 
nonradiative shock and the destruction of dust are briefly discussed in 
7.3.

\subsection{Dust destruction in interstellar shock for C20 model}

\subsubsection{Time evolution of temperature and density of gas in shock}

Figure 3 shows the structures of density (upper panel) and temperature
of gas (lower panel) in nonradiative shock at given times, and the solid 
lines in Figure 4 give the time evolution of the shock velocity 
$V_{\rm shock}$ and the gas temperature $T_{\rm shock}$ at the shock 
front.
As shown in Figure 3, the gas density increases to 4 times that in the
ISM and the gas temperature steeply raises up at the shock front which
is indicated by the arrow for a given time step.
With the initial shock velocity $\simeq$ 6000 km s$^{-1}$, the gas
temperature at the shock front remains above 10$^8$ K until 
$t \simeq 200$ years, and both $V_{\rm shock}$ and $T_{\rm shock}$ 
decrease with time.
At $t > 5 \times 10^4$ years after the explosion, the density of gas in 
the inner postshock region is about one order of magnitude lower than
that in the unshocked region, and the low-density hot bubble with
temperature of several times 10$^6$ K is formed (Fig. 3).
In this model, $V_{\rm shock}$ decelerates below 100 km s$^{-1}$ at 
$t_{\rm tr} \simeq 10^5$ years when the shock front travels the distance
of $\simeq 30$ pc and $T_{\rm shock}$ drops down to several 10$^4$ K.
After then, a dense thin-shell forms at the shock front because of the 
effective line cooling, and the SNR enters into radiative phase.

\subsubsection{Dynamics and destruction of dust grains in shock}

The motion of dust grains (upper panel) and the evolution of their sizes 
(lower panel) in postshock flow as a function of time are depicted in 
Figure 5, as an example for C grains with the initial size of 0.01
$\mu$m (dashed), 0.1 $\mu$m (dotted), and 1 $\mu$m (solid).
The thick solid curve in the upper panel indicates the position of the 
shock front. 
As dust grains initially at rest in the ISM intrude into the blast wave,
they depict the different trajectories depending on their initial sizes,
which clearly demonstrates that the dust grains are segregated and
subjected to the different sputtering processes in the postshock flow;
small grains with initial radius of 0.01 $\mu$m are quickly decelerated
by drag force, are trapped into gas near the shock front, and are 
completely destroyed by thermal sputtering.
The 0.1 $\mu$m-sized grains are gradually decelerated to comove with gas
in $7 \times 10^3$ years after entering into the shock.
Large-sized grains with radius of 1 $\mu$m are not so much suffered from
deceleration by drag force and continue to keep the high velocity
relative to gas. 
As a result, they are subjected to non-thermal sputtering, but their 
sizes are reduced very little, partly because they stay in the inner 
region of postshock flow where the gas density is lower than that near 
the shock front (see Fig. 3).

The modification of size distribution of each dust species due to
destruction by sputtering is illustrated in Figure 6; Figure 6a
for the initial size distribution before destruction and Figure 6b for
the size distribution after destruction. 
Since the erosion rate by sputtering does not strongly depend on grain 
size (see 6.2), small grains are predominantly destroyed regardless of 
grain species; the number of small-sized grain such as Al$_2$O$_3$ is
greatly reduced.
C, SiO$_2$ and Fe grains whose initial size distributions are lognormal 
with the relatively large average size are eroded but not completely 
destructed, and increase the numbers of smaller ones.
Grains with the size larger than 0.1 $\mu$m are little affected by
erosion. 
Note that the size distribution summed up over all grains species gets 
flatter for the smaller radius with time compared with the corresponding
initial size distribution approximated by a power-law formula with index
of $-2.5$, while that for the larger radius remains almost unchanged.

\subsubsection{Efficiency of dust destruction}

The destruction efficiency $\epsilon_j$ of each grain species is given
in Table 5 (see unmixed grain model) along with the initial average 
radius.
It is expected that the efficiency of dust destruction are higher for 
grain species with the smaller average size because the smaller grains 
are predominantly destructed by sputtering.
In fact, the efficiency of destruction of Al$_2$O$_3$ grain with the 
smallest average radius is 0.667 and is the highest among dust species 
considered here.
Si and Fe grains with initial average radii larger than 0.1 $\mu$m have 
the destruction efficiencies less than 0.2; 0.13 for Si grain and 0.15
for Fe grain.
However, the destruction efficiency of C grain with the smaller average 
radius is smaller than that of SiO$_2$ grain because of the lowest 
erosion rate among all grain species at $T \ga 2 \times 10^6$ K (see 
Fig. 2).
Also, FeS grain has higher destruction efficiency (0.578) than those of 
Mg$_2$SiO$_4$ (0.451) and MgO grains (0.505) despite of the larger 
initial average radius, which reflects not only the higher erosion rate
at $T \ga 10^7$ K but also the mass distribution much more weighted
toward the smaller grains.
Likewise, the destruction efficiency of MgSiO$_3$ grain with average 
size comparable to that of Mg$_2$SiO$_4$ and MgO is high (0.637)
because of the lack of large grains.
These facts indicate that the efficiency of dust destruction depends on 
the initial size distribution of dust grains as well as on the
sputtering yield.

In order to clarify the effect of the initial size distribution on the 
efficiency of dust destruction, we present the results of dust
destruction for the mixed grain model calculated with the same SN model 
and value of $n_{{\rm H}, 0}$ as those for the unmixed grain model.   
Figure 7 shows the size distribution of each grain species (Al$_2$O$_3$,
MgSiO$_3$, Mg$_2$SiO$_4$, SiO$_2$ and Fe$_3$O$_4$) before destruction 
(Fig. 7a) and after destruction (Fig. 7b), and the destruction 
efficiencies of these grains are tabulated in Table 5 (see mixed
grain model).
As is the same as the unmixed grain model, the numbers of Al$_2$O$_3$ 
and Fe$_3$O$_4$ grains with the small average size ($<$ a few tens \AA) 
are considerably reduced, which results in high efficiencies of
destruction ($> 0.7$).
For MgSiO$_3$, Mg$_2$SiO$_4$ and SiO$_2$ grains with lognormal size 
distributions, the erosion of large grains leads to the increase of the 
number of smaller ones, and their destruction efficiencies span the
range of $0.4 \la \epsilon_j \la 0.64$.
Note that although the average size is twice larger than that in the 
unmixed case, the destruction efficiency (0.59) of Mg$_2$SiO$_4$ grain
in the mixed case is significantly larger than that (0.45) in the 
unmixed case.
The reason is that Mg$_2$SiO$_4$ grains formed in the unmixed ejecta
have power-law-like size distribution and include the large-sized grains 
of $> 0.1$ $\mu$m almost undestructed in shock; the average size is not
always suitable for assessing the feasibility of dust destruction.
Thus, we conclude that the dust destruction efficiency is very sensitive 
to the initial size distribution.
The mass fraction of dust destructed reaches up to 34 \% for the unmixed 
grain model and 50 \% for the mixed grain model.

\subsection{Time-scale of dust destruction in the early universe}

In this subsection, we investigate the dependences of the efficiency of 
dust destruction $\epsilon_j$ and the mass of gas swept up by shock 
$M_{\rm swept}$ on $E_{51}$, $M_{\rm pr}$ and $n_{\rm H, 0}$, and derive 
an analytic formula describing the time-scale of dust destruction 
for the unmixed grain model.
Figure 8 shows the destruction efficiency of each grain species versus
SN explosion energy calculated for $n_{\rm H, 0}=1$ cm$^{-3}$; 
Figure 8a is for Al$_2$O$_3$, FeS, Mg$_2$SiO$_4$ and Fe grains, and 
Figure 8b is for MgSiO$_3$, MgO, SiO$_2$, C and Si grains.
Also, the overall efficiency of dust destruction, which is defined as
the ratio of the total mass of dust destructed to the total mass of dust 
swept up by shock, is plotted in Figure 8a.
The SN models used for the calculations are distinguished by open 
circles (CCSNe), open squares (HNe) and filled triangles (PISNe).

The destruction efficiencies for each grain species are almost the same
for CCSNe and a PISN with the explosion energy of $E_{51}=1$
irrespective of the progenitor mass $M_{\rm pr}$, and increase with 
increasing $E_{51}$. 
Note that high explosion energy with $E_{51} \ge 10$ causes the 
temperature of gas in postshock flow to raise up as high as $10^9$ K,
but this does not directly influence on the efficiency of dust
destruction because the increase of gas temperature does not always lead 
to the enhancement of the erosion rate by sputtering (see Fig. 2a).
The reason for the increased efficiency with increasing $E_{51}$ is 
considered as follows. 
The high-velocity shock (the initial shock velocity $\ga 10^4$ km
s$^{-1}$) generated by the energetic SN explosion induces high velocity 
of dust relative to gas. 
Then, dust grains are efficiently decelerated by drag force (see 
Equation (19)), are trapped in the high-density region near the shock 
front, and are significantly eroded by thermal sputtering. 
Furthermore, since the high shock velocity takes much longer time to
drop down to 100 km s$^{-1}$ (see Fig. 9), dust grains are immersed in a 
hot plasma for a long time, which causes even larger grains to be more 
eroded.

It should be pointed out here that the efficiencies of destruction for 
the models of P170 with $E_{\rm 51} =20$ and P200 with $E_{\rm 51} =28$ 
are a little higher than that for H30 with $E_{\rm 51} =30$.  
The reason is considered that the ejecta mass of PISNe is more than 6 
times larger than that of HNe, and the resulting longer duration of the 
free expansion phase causes the longer truncation time than HNe
(Fig. 9). 
On the other hand, the efficiency of destruction for P150 with 
$M_{\rm pr}=150$ $M_\odot$ and $E_{\rm 51} =1$ is almost the same as
that for CCSNe with $E_{\rm 51} =1$ because the initial shock velocity 
($\sim$ 3000 km s$^{-1}$) much lower than that of CCSNe makes the
truncation time comparable to CCSNe. 
Therefore, although the destruction efficiency $\epsilon_j$ is almost 
independent of $M_{\rm pr}$ as long as $E_{51}=1$,  $\epsilon_j$ for 
PISNe higher than that for HNe at $E_{51} \ge 10$ reflects the 
difference in the explosion mechanism depending on the progenitor mass. 

To examine the dependence of $\epsilon_j$ on $E_{51}$ for each type of
SN, we calculate the coefficients $a_{1,j}$ and $b_{1,j}$ in Equation
(A1) given in Appendix for SNe II (C13, C20, C25, C30, H25 and H30) and 
PISNe (P150, P170 and P200) separately, and the calculated overall 
efficiencies of dust destruction are indicated by dotted (SNe II) and 
dashed lines (PISNe) in Figure 8a.
Although the difference in the efficiency of dust destruction between
SNe II and PISNe increases with increasing $E_{51}$, the deviations
from the values calculated by $a_{1,j}$ and $b_{1,j}$ for all SN models 
are at most about 10 \% at $E_{51}=30$. 
Thus, we consider that $\epsilon_j$ is almost independent of $M_{\rm
pr}$ as long as $E_{\rm 51}$ is the same. 
For reference, we tabulate the values of $a_{1,j}$ and $b_{1,j}$ for 
SNe II and PISNe in Table 6. 
The mass of gas swept up by shock $M_{\rm swept}$ and the truncation
time $t_{\rm tr}$ calculated for $n_{\rm H, 0}=1$ cm$^{-3}$ are
presented in Figure 9 as a function of SN explosion energy.
As is the same as $\epsilon_j$, $M_{\rm swept}$ and $t_{\rm tr}$ are 
almost the same for $E_{51}=1$ regardless of $M_{\rm pr}$, and increase 
with increasing $E_{51}$.

Next, we show the dependences of $\epsilon_j$ and $M_{\rm swept}$ on
the preshock gas density $n_{\rm H, 0}$.
Figure 10a plots the overall efficiency of dust destruction versus 
$n_{\rm H, 0}$ for C20 (crosses), H25 (open circles) and H30 models
(filled triangles) with $E_{51}= 1, 10$ and 30, respectively.
The efficiency of dust destruction increases with increasing $n_{\rm H,
0}$, since higher gas density causes more frequently collisions between 
dust and gas to erode efficiently the surface of dust grains by 
sputtering; for example, the mass fraction of dust destroyed for H30 
model is 78 \% for $n_{\rm H, 0} =10$ cm$^{-3}$, but only 23 \% for 
$n_{\rm H, 0} =0.1$ cm$^{-3}$.
In Figure 10b, we present the mass of gas swept up by shock 
$M_{\rm swept}$ as a function of $n_{\rm H,0}$ for C20, H25 and H30
models. 
Note that $M_{\rm swept}$ decreases with increasing $n_{\rm H, 0}$,
because shock wave more quickly decelerates and travels only small
distance.

The approximation formulae presented in Appendix being combined, the
time-scale of destruction for each dust species by the interstellar
shock in the early universe is presented by
\begin{eqnarray}
\tau^{-1}_{{\rm SN}, j}= \epsilon_j(E_{51}, n_{\rm H,0})\frac{4144 E_{51}^{0.8}
n_{\rm H,0}^{-0.142 E_{51}^{0.063}} M_\odot}{M_{\rm ISM}} \gamma_{\rm SN}, 
\end{eqnarray}
for all SN models, where the dependences of $\epsilon_j$ on $E_{51}$ and 
$n_{\rm H,0}$ are given by Equations (A1) and (A3), respectively.
We derived the time-scale of dust destruction as a function of not only 
the explosion energy of SN but also the gas density in the ISM.
The swept-up gas mass is proportional not to $E_{51}$ but to 
$\sim E_{51}^{0.8}$, being different from  the formula proposed by 
McKee (1989). 
The time-scale of dust destruction derived here could be applicable to 
investigate the time evolution of dust grains in the early universe. 
Especially, the difference of the efficiency of dust destruction for
each grain species may have great influences on the amount and size 
distribution of dust grains residing in the early interstellar space.

\subsection{The effects of the cooling processes of gas on dust 
destruction}

Figure 11 shows the cumulative energy lost by the atomic process 
(dashed), the inverse Compton scattering (dotted) and the thermal 
emission from dust (solid) calculated for the SN model of C20 and for 
the ISM with parameters of $n_{{\rm H}, 0} = 1$ and $Z = 10^{-4}$ 
$Z_\odot$ at the redshift of $z=20$.
The total energy lost by these cooling processes is drawn by the thick 
solid curve.
The inverse Compton cooling is comparable to that by the atomic process 
in the early phase of the SNR ($t < 10^4$ years).
As the gas temperature decreases, the atomic process by H and He line 
coolings becomes dominant. 
Compared with the above two cooling processes, the thermal emission from 
dust is extremely low and contributes only less than 0.1 \% to the total 
energy loss. 
The transition of nonradiative shock to radiative shock occurs when the 
cumulative energy loss reaches to 0.01 \% of the explosion energy. 
Thus, only the H and He line coolings affect the time evolution of  
nonradiative shock and the dust destruction efficiencies in the ISM with 
metallicity less than $Z = 10^{-4}$ $Z_\odot$ corresponding to 
the dust-to-gas mass ratio of $4.5 \times 10^{-7}$ in this paper.  

However, it is expected that the effects of cooling of gas by dust
come to be important for the ISM with higher dust-to-gas mass ratio.
The result of the calculation for dust destruction in the ISM with 
$Z = 10^{-4}$ $Z_\odot$ and the dust-to-gas mass ratio $4.5 \times
10^{-3}$ shows that the truncation time and the overall efficiency of
dust destruction decrease by 4.4 \% and 2.6 \%, respectively, compared 
with the results for dust-to-gas mass ratio of $4.5 \times 10^{-7}$, 
though the gas temperature in postshock gas decreases by about 20\%.  
Even if the dust-to-gas mass ratio is raised up to by four orders of 
magnitude, the cooling of gas by dust in nonradiative shock does not 
have the significant effects on the efficiency of dust destruction as 
well as the evolution of nonradiative shock. 

Furthermore, the inverse Compton cooling has influences on the evolution 
of nonradiative shock in the early universe.
Because the cooling rate by the inverse Compton scattering is 
proportional to $n_e T (1+z)^4$ and that by atomic process is 
proportional to $n_e n_{\rm H} \sim n_{\rm H}^2$, the contribution 
of the inverse Compton cooling is enhanced in postshock flow with the 
lower gas density and/or the higher gas temperature at the higher
redshift.
At the redshift of $z=40$, for the SN model C20 with the preshock gas
density of $n_{{\rm H}, 0}=1$ cm$^{-3}$, the truncation time is only 2
\% shorter than that calculated at $z = 20$.
Thus, in this case, the inverse Compton cooling seems not to affect the 
destruction efficiency of dust grains. 
However, more systematic studies covering wide ranges of the SN
explosion energy and the gas density in the ISM are necessary to reveal 
the effect of the inverse Compton cooling.

Note that the cooling function of gas for $Z=10^{-3}$ $ Z_{\odot}$ is 
almost the same as that for the zero metal case (Sutherland \& 
Dopita 1993). 
Thus, the parameters of the ISM considered in this paper ($Z = 10^{-4}$ 
$Z_\odot$ and $z=20$) is competent for investigating dust destruction in 
the early universe, and the time-scale of dust destruction derived in
7.2 is applicable over the wide ranges of metallicity of gas in the ISM 
($Z \la 10^{-3}$ $Z_\odot$) and redshift ($z \le 40$).

\section{Summary}

We investigate the destruction of dust grains in the interstellar shocks 
driven by SNe as the second step to reveal the evolution of dust in the 
early universe, based on the dust models obtained by Nozawa et
al. (2003).
We focus on dust destruction in nonradiative shock where dust grains are 
predominantly destroyed by non-thermal and thermal sputtering because of
high temperature and high velocity of gas.
The sputtering yields for the combinations of dust and ion species of 
interest to us are evaluated by applying the universal relation for 
sputtering yield with a slight modification and by determining the
value of $K$ by fitting to the available experimental data and/or the 
results of the EDDY simulations.
The modified universal relation derived here can present the better fits 
to the available experimental data than the previous one.

In the calculations of dust destruction, we solve the time evolution of 
gas temperature and density in spherically symmetric shocks, adopting 
the hydrodynamical models by Umeda \& Nomoto (2002) as the initial 
condition for interstellar shock and including the cooling of gas by 
the atomic process, the inverse Compton scattering and the thermal 
emission of dust.
The erosion of dust by thermal and non-thermal sputtering caused by 
motion of dust relative to gas is carefully treated by taking into account 
the size distribution of each dust species.
The results of calculations are summarized as follows.

 (1) Because the sputtering predominantly destroys the small grain, the 
 number of small-sized grain such as Al$_2$O$_3$ and MgSiO$_3$ is 
 greatly reduced.
 The erosion of C, SiO$_2$ and Fe grains whose size distributions are
 lognormal-like with the average size larger than $0.01$ $\mu$m
 increases the numbers of smaller ones. 
 The size distribution summed up over all grains species becomes flatter 
 for the small radius compared with the initial size distribution, while 
 that for the radius larger than 0.2 $\mu$m remains almost unchanged.

 (2) The efficiency of dust destruction is higher for the grains with 
 the small average size such as Al$_2$O$_3$ and MgSiO$_3$ and with the 
 power-law-like size distribution such as FeS, Mg$_2$SiO$_4$ and MgO.
 On the other hand, Si, Fe, C and SiO$_2$ grains, which have the
 lognormal-like size distribution and the relatively large average 
 radius, have lower efficiency of destruction. The detailed analysis of
 the behavior of dust destruction efficiency for each grain species
 indicates that not only sputtering yields but also the initial size 
 distribution plays a crutial role in the efficiency of dust destruction 
 by sputtering.  

 (3) The efficiency of destruction $\epsilon_j$ for each dust species
 increases with increasing the explosion energy $E_{51}$ and/or with 
 increasing the preshock gas density $n_{\rm H,0}$, but is almost 
 independent of the progenitor mass $M_{\rm pr}$ of SN as long as 
 $E_{51}$ is the same. 
 The destruction efficiency $\epsilon_j$ as a function of $E_{51}$ is 
 well reproduced by a power-law formula given by Equation (A1).
 The dependence of $\epsilon_j$ on $n_{\rm H,0}$ is well expressed by 
 the quadratic equation (A3) in terms of $\log(n_{\rm H,0})$.

 (4) As is the same as $\epsilon_j$, the mass of gas swept up by shock 
 wave $M_{\rm swept}$ is the increasing function of $E_{51}$ and is 
 approximated by a power-law formula given by Equation (A2).
 However, $M_{\rm swept}$ decreases with increasing the gas density in
 the ISM, and its dependence is also reproduced by a power-law formula 
 whose index is given by Equation (A4) as a function of $E_{51}$.
 Finally, by combining these results, we present the analytic formula 
 for the time-scale of destruction for each grain species in the 
 early universe as a function of $E_{51}$ and $n_{\rm H,0}$.
 The derived time-scale of dust destruction can be employed 
 to investigate the time evolution of the amount of dust grains in the 
 early universe. 

 (5) In the early universe, only the H and He line coolings affect the 
 time evolution of nonradiative shock and the efficiencies of dust destruction.
 The thermal emission from dust grains is not so much important even for 
 the ISM with the dust-to-gas mass ratio of $\sim 10^{-3}$ as long as the 
 metallicity in the ISM is $Z \la 10^{-3}$ $Z_\odot$.
 Also, it is expected that the inverse Compton cooling does not have the 
 great effects on the evolution of shock at high redshifts.
 Thus, the time-scale of dust destruction derived in this paper is 
 applicable for ranges of metallicity of gas in the ISM 
 of $Z \la 10^{-3}$ $Z_\odot$ and redshift of $ z \le 40$.

In this study, we focused on dust destruction in nonradiative shock
where dust grains in the ISM are considered to be predominantly 
destroyed.
We should be mentioned here that a part of dust grains formed in the 
ejecta of SNe are 
expected to be destroyed by reverse shock penetrating into the ejecta.
Because sputtering is the most dominant destruction process of dust
grains in reverse shock as well, the simulation code of dust destruction 
constructed in this paper can be applied to explore the destruction of 
dust grains by reverse shocks.
This work is now in progress.
Furthermore, in radiative shock with the shock velocities below 100 km 
s$^{-1}$, shattering and partial vaporization by grain-grain collisions 
become important in the presence of a magnetic field. 
A magnetic field frozen into the gas causes charged grains to gyrate 
and accelerate behind the shock front and results in the enhancement of 
not only the sputtering rate but also grain-grain collision frequency.
The shattering by grain-grain collision mainly leads to the 
redistribution of grain size (Tielens et al. 1994; Jones, Tielens, \& 
Hollenbach 1996).
The sophisticated study on this subject is left for the future work.

\acknowledgments

The authors are grateful to the anonymous referee for critical comments 
which have improved the manuscript.
The authors thank Dr. H. Umeda and Prof. K. Nomoto for making the ejecta
model of Pop III supernovae available.
This work has been supported in part by a Grant-in-Aid for Scientific
Research from the Japan Society for the Promotion of Sciences (13640229).

\appendix

\section{The appriximation formulae for $\epsilon_j$ and $M_{\rm swept}$}

In this Appendix,  
we present the approximate formulae for the efficiency of
dust destruction $\epsilon_j$ and the mass of gas swept up by shock 
$M_{\rm swept}$ as function of the SN explosion energy $E_{51}$ and the 
gas density $n_{{\rm H} ,0}$ in the ISM.
The energy dependence of the destruction efficiency for each grain 
species can be well reproduced by a power-law formula given by 
\begin{eqnarray}
\epsilon_j = a_{1,j} E_{51}^{b_{1,j}}.
\end{eqnarray}
The solid lines in Figure 8 indicate the least-squares fittings by this 
power-law formula for all SNe whose numerical coefficients 
$a_{1,j}$ and $b_{1,j}$ are tabulated in Table 6.
The dependence of $M_{\rm swept}$ on $E_{51}$ is well fitted by a 
power-law formula
\begin{eqnarray}
M_{\rm swept} = a_2 E_{51}^{b_2}
\end{eqnarray}
using the coefficients $a_2$ and $b_2$ given in Table 6, and
is depicted in Figure 9 for all SNe (solid), SNe II (dotted), and 
PISNe (dashed).
The destruction efficiency $\epsilon_j$ for each grains species $j$ as
well as the overall destruction efficiency as a function of $n_{\rm
H,0}$ is expressed by  
\begin{eqnarray}
\log{\epsilon_j} = c_{j} \left(\log{n_{\rm H,0}}\right)^2 + d_{j}
\log{n_{\rm H,0}} + e_{j},
\end{eqnarray}
and the coefficients $c_{j}$, $d_{j}$ and $e_{j}$ derived by the least 
squares fittings are given in Table 7 for the models C20, H25 and H30, 
respectively, along with the coefficients for overall destruction 
efficiency.  
The solid curves in Figure 10a depict the overall destruction 
efficiencies calculated by Equation (A3) for the models C20, H25 and 
H30.
This formula reproduces the destruction efficiencies calculated by the 
simulations with the accuracy less than 10 \% for $n_{\rm H,0} \le 10$ 
cm$^{-3}$. 
The calculated $M_{\rm swept}$ can be reproduced by a power-law formula 
$M_{\rm swept} \propto n_{\rm H, 0}^g $ where the index $g$ is weakly 
dependent on $E_{51}$ and is approximated by
\begin{eqnarray}
g = -0.142 E_{51}^{0.063}.
\end{eqnarray}
This formula gives good agreements with the results of simulations,
as represented by the solid lines in Figure 10b.

\clearpage

\begin{deluxetable}{lccccc}
\tablecaption{Elemental Abundances of Gas in the ISM 
\label{tbl-2}}
\tablewidth{0pt}
\tablehead{
\colhead{$Z/Z_\odot$}& \colhead{$A_{\rm He}$} & \colhead{$A_{\rm C}$}
& \colhead{$A_{\rm N}$} & \colhead{$A_{\rm O}$} & \colhead{$A_{\rm e}$}}
\startdata
$10^{-4}$ & $8.04 \times 10^{-2}$ & $3.62 \times 10^{-8}$ 
& $1.12 \times 10^{-8}$ & $2.70 \times 10^{-7}$ & 1.128 \\
% 1        & $9.75 \times 10^{-2}$ & $3.62 \times 10^{-4}$
%& $1.12 \times 10^{-4}$ & $8.53 \times 10^{-4}$ & 1.120 \\
\enddata
\tablecomments{The number abundance $A_i$ relative to that of hydrogen 
atom in the ISM. Note that the abundances of C, N and O are
 calculated by multiplying the metallicity of $Z=10^{-4}$ $Z_{\odot}$ and
 the primordial ratios in Table 4 of Sutherland \& Dopita (1993).}
\end{deluxetable}

\clearpage

\begin{deluxetable}{lccccc}
\tablewidth{0pt}
\tablecaption{Dust Models Used in the Calculations}
\tablehead{ 
& & \multicolumn{2}{c}{unmixed grain model} 
& \multicolumn{2}{c}{mixed grain model} \\
\colhead{dust species} & \colhead{$\rho_{{\rm d}, j}$ (g cm$^{-3}$)} &
\colhead{$A_j$} & \colhead{$a_{{\rm ave}, j}$ ($\mu$m)} & 
\colhead{$A_j$} & \colhead{$a_{{\rm ave}, j}$ ($\mu$m)}
}
\startdata
C             & 2.26 & 0.103 & 0.029  & &  \\
Si            & 2.32 & 0.304 & 0.250  & &  \\
Fe            & 7.89 & 0.050 & 0.148  & &  \\
FeS           & 4.84 & 0.156 & 0.0088 & &  \\
Al$_2$O$_3$   & 3.98 & 0.002 & 0.0007 & 0.001 & 0.0003 \\
MgSiO$_3$     & 3.18 & 0.010 & 0.0045 & 0.102 & 0.011 \\
Mg$_2$SiO$_4$ & 3.20 & 0.219 & 0.0066 & 0.205 & 0.010 \\
SiO$_2$       & 2.64 & 0.100 & 0.047  & 0.588 & 0.033 \\
MgO           & 3.56 & 0.056 & 0.0058 & &  \\
Fe$_3$O$_4$   & 5.21 &       &        & 0.103 & 0.0022 \\
\enddata
\tablecomments{The models of dust grains are adopted from the results of 
the calculations of dust formation in the ejecta of Pop III SNe with 
$M_{\rm pr} = 20$ $M_\odot$ by Nozawa et al. (2003).
The bulk density of grain species $j$ is denoted by $\rho_{{\rm d}, j}$. 
The mass fraction $A_j$ and the average size $a_{{\rm ave},j}$ of each 
grain species are given for the unmixed and mixed grain models.
Note that the average radius of Si grain in the unmixed grain model is
calculated for the grains larger than 0.03 $\mu$m which occupy 99.8 \%
 of the total mass.
}
\end{deluxetable}

\clearpage

\begin{deluxetable}{lcc}
\tablecaption{Models of Supernova Ejecta 
\label{tbl-2}}
\tablewidth{0pt}
\tablehead{
& \colhead{Progenitor mass} & \colhead{Explosion energy} \\
\colhead{Model} & \colhead{($M_{\odot}$)} & \colhead{$E_{51}$ 
(10$^{51}$ ergs)}}
\startdata
C13  & 13  & 1 \\
C20  & 20  & 1 \\
C25  & 25  & 1 \\
C30  & 30  & 1 \\
H25  & 25  & 10 \\
H30  & 30  & 30 \\
P150 & 150 & 1 \\
P170 & 170 & 20 \\
P200 & 200 & 28 \\
\enddata
\tablecomments{The models of the SN ejecta are adopted from Umeda \&
Nomoto (2002). 
In the model names, the labels C, H and P represent CCSNe, HNe
and PISNe, respectively, and the numerical value denotes the mass of the 
progenitor in units of solar mass.}
\end{deluxetable}

\clearpage

\begin{deluxetable}{lccccl}
\tablewidth{0pt}
\tablecaption{Parameters for Calculations of Sputtering Yield and 
References for Yield Data}
\tablehead{
\colhead{dust species} & \colhead{$U_0$ (eV)}   &
\colhead{$Z_{\rm d}$}  & \colhead{$M_{\rm d}$}  &
\colhead{$K$}          & \colhead{references}
}
\startdata
C & 4.0 & 6 & 12 & 0.61 & 1, 2, 3, 4, 5 \\
Si & 4.66 & 14 & 28 & 0.43 & 3, 4, 6, 7, 8, 9, 10, 11\\
Fe & 4.31 & 26 & 56 & 0.23 & 3, 4, 5, 7, 8, 12, 13 \\
FeS & 4.12 & 21 & 44 & 0.18 & EDDY \\
Al$_2$O$_3$ & 8.5 & 10 & 20.4 & 0.08 & 3, 14, 15 \\
MgSiO$_3$ & 6.0 & 10 & 20 & 0.1 & 16 \\
Mg$_2$SiO$_4$ & 5.8 & 10 & 20 & 0.1 & 16 \\
SiO$_2$ & 6.42 & 10 & 20 & 0.1 &  3, 14, 15, 17, 18 \\
MgO & 5.17 & 10 & 20 & 0.06 & 14, 15, EDDY \\
Fe$_3$O$_4$ & 4.98 & 15.7 & 33.1 & 0.15 & EDDY  \\
\enddata
\tablecomments{The surface binding energy is represented by $U_0$ in
 units of eV. 
The mean atomic number and mean mass number of the target material are
$Z_{\rm d}$ and $M_{\rm d}$, respectively, and the free parameter $K$ is 
determined by fitting to the available experimental data and/or the 
results of the EDDY simulation. See text for details.}
\tablerefs{
(1) Bohdansky, Bay, \& Ottenberger (1978); 
(2) Bohdansky, Roth, \& Sinha (1976);
(3) Roth, Bohdansky, \& Ottenberger (1979); 
(4) Rosenberg \& Wehner (1962);
(5) Hechtl, Bohdansky, \& Roth (1981); 
(6) Laegreid \& Wehner (1961);
(7) Southern, Willis, \& Robinson (1963);
(8) Eernisse (1971);
(9) Sommerfeldt, Mashkova, \& Molchanov (1972);
(10) Coburn, Winters, \& Chuang (1977);
(11) Blank \& Wittmaack (1979);
(12) von Seefeld et al. (1976);
(13) Bohdansky, Bay, \& Roth (1977);
(14) Bach (1970);
(15) Nenadovic et al. (1990);
(16) Tielens et al. (1994);
(17) Bach, Kitzmann, \& Schroder (1974);
(18) Edwin (1973)
}
\end{deluxetable}

\clearpage

\begin{deluxetable}{lcccc}
\tablewidth{0pt}
\tablecaption{The Efficiency of Dust Destruction for C20 Model with 
$n_{{\rm H}, 0} = 1$}
\tablehead{
\colhead{} & \multicolumn{2}{c}{unmixed grain model} 
& \multicolumn{2}{c}{mixed grain model} \\
\colhead{} & \colhead{$\epsilon_j$} & \colhead{$a_{{\rm ave}, j}$} ($\mu$m)
& \colhead{$\epsilon_j$} & \colhead{$a_{{\rm ave}, j}$} ($\mu$m)}
\startdata
C             & 0.247 & 0.029  & & \\ 
Si            & 0.134 & 0.250  & & \\
Fe            & 0.154 & 0.148  & & \\ 
FeS           & 0.578 & 0.0088 & & \\
Al$_2$O$_3$   & 0.667 & 0.0007 & 0.794 & 0.0003 \\
MgSiO$_3$     & 0.637 & 0.0045 & 0.631 & 0.011  \\
Mg$_2$SiO$_4$ & 0.451 & 0.0066 & 0.586 & 0.010  \\
SiO$_2$       & 0.411 & 0.047  & 0.399 & 0.033  \\
MgO           & 0.505 & 0.0058 & &   \\
Fe$_3$O$_4$   & & & 0.741 & 0.0022 \\
overall       & 0.340 & & 0.497 & \\
\enddata
\end{deluxetable}

\clearpage

\begin{deluxetable}{lcccccc}
\tablewidth{0pt}
\tablecaption{Coefficients for the Approximation Formula (A1) and (A2)}
\tablehead{
\colhead{}    &  \multicolumn{2}{c}{for all SNe} & 
\multicolumn{2}{c}{for SNe II} & 
\multicolumn{2}{c}{for PISNe} \\
\colhead{dust species} & 
\colhead{$a_{1, j}$}   & \colhead{$b_{1, j}$} & 
\colhead{$a_{1, j}$}   & \colhead{$b_{1, j}$} & 
\colhead{$a_{1, j}$}   & \colhead{$b_{1, j}$} 
}
\startdata
C & 2.532E-01 & 1.660E-01 & 2.526E-01 & 1.407E-01 & 2.580E-01 &
 1.820E-01 \\
Si & 1.380E-01 & 2.371E-01 & 1.377E-01 & 1.979E-01 & 1.414E-01 &
 2.631E-01 \\
Fe & 1.583E-01 & 2.402E-01 & 1.586E-01 & 2.005E-01 & 1.594E-01 &
 2.707E-01 \\
FeS & 5.842E-01 & 7.897E-02 & 5.850E-01 & 6.758E-02 & 5.834E-01 &
 8.857E-02 \\
Al$_2$O$_3$ & 6.724E-01 & 5.649E-02 & 6.733E-01 & 4.803E-02 & 6.709E-01 &
 6.393E-02 \\
MgSiO$_3$ & 6.429E-01 & 6.358E-02 & 6.438E-01 & 5.421E-02 & 6.417E-01 &
 7.165E-02 \\
Mg$_2$SiO$_4$ & 4.585E-01 & 1.146E-01 & 4.590E-01 & 9.827E-02 & 4.595E-01 &
 1.275E-01 \\
SiO$_2$ & 4.183E-01 & 1.238E-01 & 4.189E-01 & 1.054E-01 & 4.19E-01 &
 1.383E-01 \\
MgO & 5.117E-01 & 1.046E-01 & 5.122E-01 & 9.020E-02 & 5.124E-01 &
 1.160E-01 \\
overall & 3.458E-01 & 1.307E-01 & 3.460E-01 & 1.099E-01 & 3.476E-01 &
 1.463E-01 \\
\tableline
 & & & & & & \\
\tableline
& $a_2$ & $b_2$ & $a_2$ & $b_2$ & $a_2$ & $b_2$ \\
\tableline
$M_{\rm swept}$ & 4.144E+03 & 7.967E-01 & 4.118E+03 & 6.627E-01 &
 4.481E+03 & 8.864E-01 \\
\enddata
\end{deluxetable}

\clearpage

\begin{deluxetable}{lccccccccc}
\tablewidth{0pt}
\tablecaption{Coefficients for the Approximation (A3)}
\tablehead{
\colhead{}    &  \multicolumn{3}{c}{C20} & 
\multicolumn{3}{c}{H25} & 
\multicolumn{3}{c}{H30} \\
\colhead{dust species} & 
\colhead{$c_{j}$} & \colhead{$d_{j}$} & \colhead{$e_{j}$} & 
\colhead{$c_{j}$} & \colhead{$d_{j}$} & \colhead{$e_{j}$} &
\colhead{$c_{j}$} & \colhead{$d_{j}$} & \colhead{$e_{j}$}
}
\startdata
C             & $-0.062$ & 0.41 & $-0.61$ & $-0.07$  & 0.36 & $-0.46$ &
 $-0.08$  & 0.33 & $-0.39$  \\
Si            & $-0.044$ & 0.54 & $-0.87$ & $-0.06$  & 0.50 & $-0.67$ &
 $-0.072$ & 0.47 & $-0.58$  \\
Fe            & $-0.057$ & 0.55 & $-0.81$ & $-0.078$ & 0.49 & $-0.61$ &
 $-0.096$ & 0.46 & $-0.51$  \\
FeS           & $-0.088$ & 0.23 & $-0.23$ & $-0.077$ & 0.18 & $-0.16$ &
 $-0.073$ & 0.16 & $-0.13$  \\
Al$_2$O$_3$   & $-0.031$ & 0.14 & $-0.17$ & $-0.030$ & 0.12 & $-0.13$ &
 $-0.032$ & 0.11 & $-0.11$  \\
MgSiO$_3$     & $-0.060$ & 0.18 & $-0.19$ & $-0.057$ & 0.14 & $-0.13$ &
 $-0.055$ & 0.12 & $-0.11$  \\
Mg$_2$SiO$_4$ & $-0.087$ & 0.31 & $-0.35$ & $-0.092$ & 0.25 & $-0.24$ &
 $-0.097$ & 0.22 & $-0.19$  \\
SiO$_2$       & $-0.086$ & 0.33 & $-0.39$ & $-0.093$ & 0.27 & $-0.27$ &
 $-0.097$ & 0.24 & $-0.22$  \\
MgO           & $-0.079$ & 0.29 & $-0.30$ & $-0.086$ & 0.23 & $-0.20$ &
 $-0.089$ & 0.20 & $-0.16$  \\
overall       & $-0.067$ & 0.33 & $-0.47$ & $-0.067$ & 0.29 & $-0.35$ &
 $-0.072$  & 0.26 & $-0.30$ \\
\enddata
\end{deluxetable}

\clearpage

\begin{figure}
\epsscale{0.8}
\plottwo{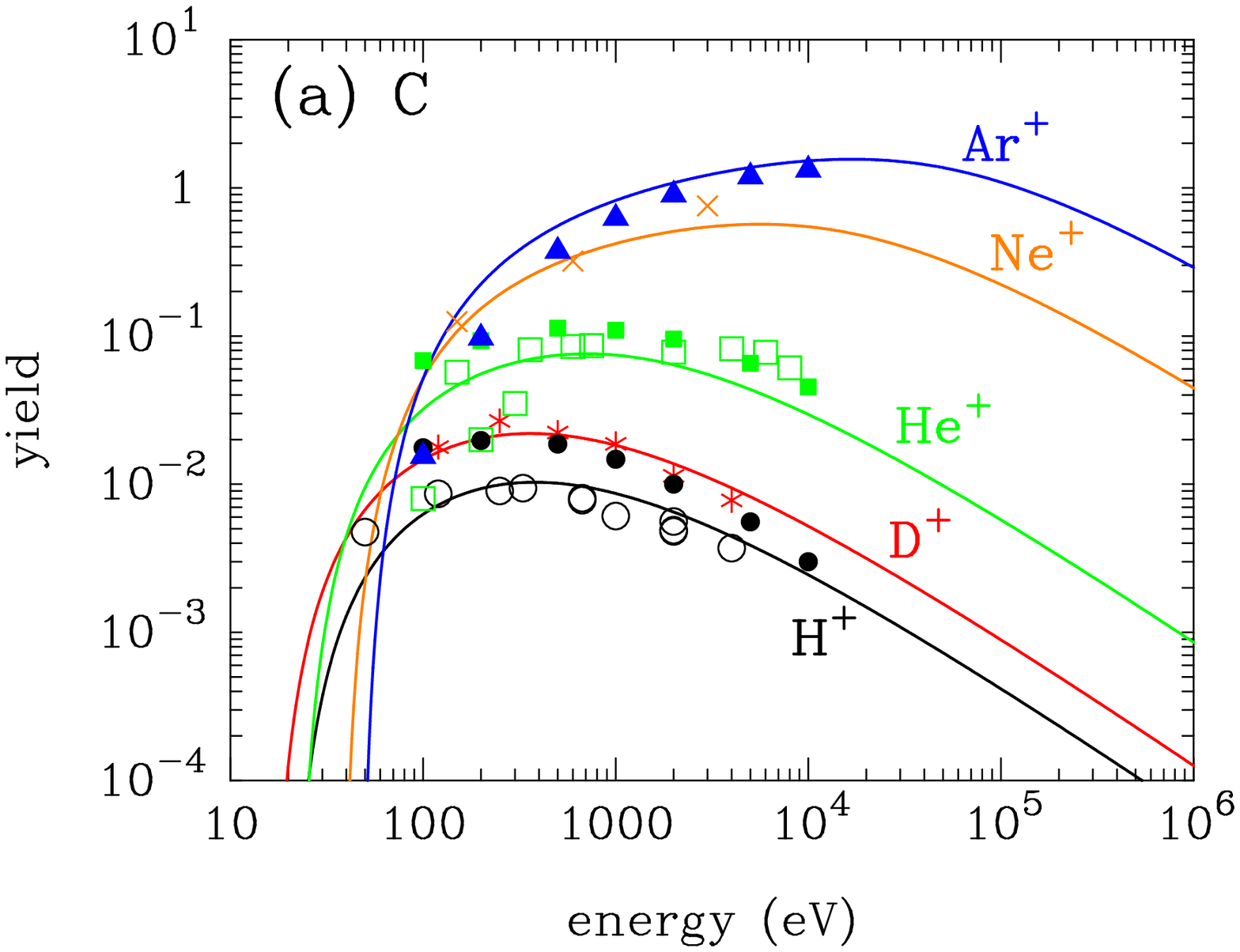}{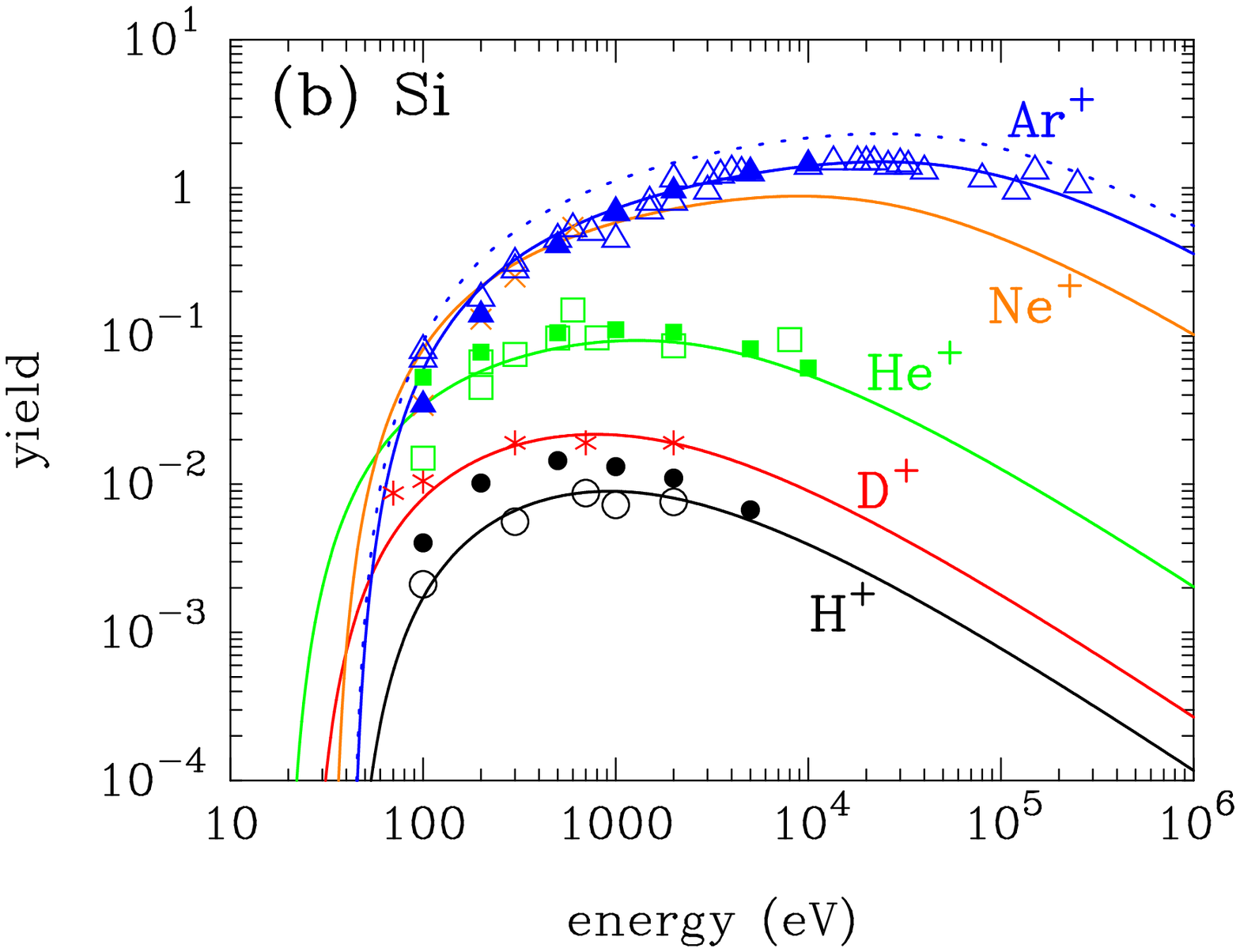}
\plottwo{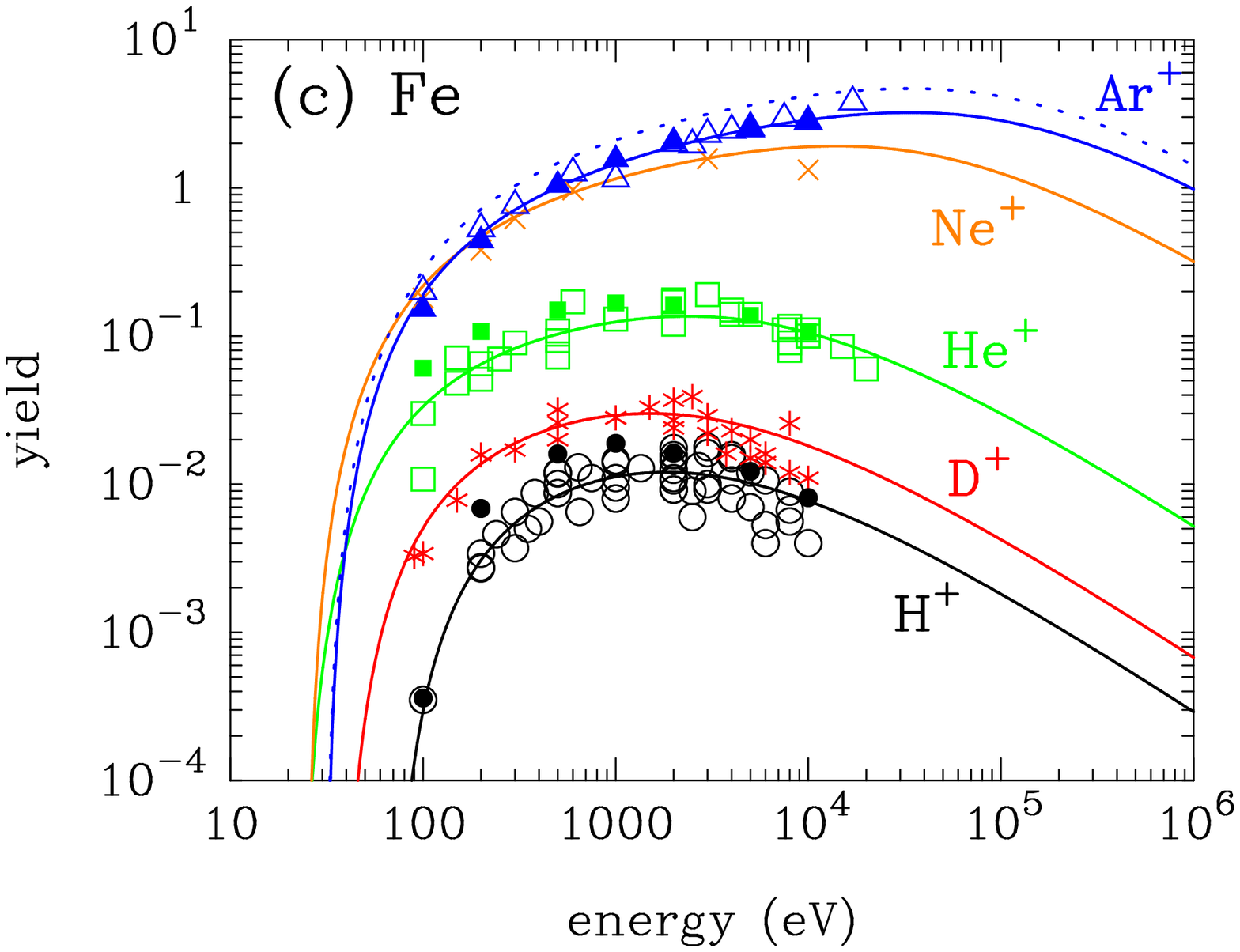}{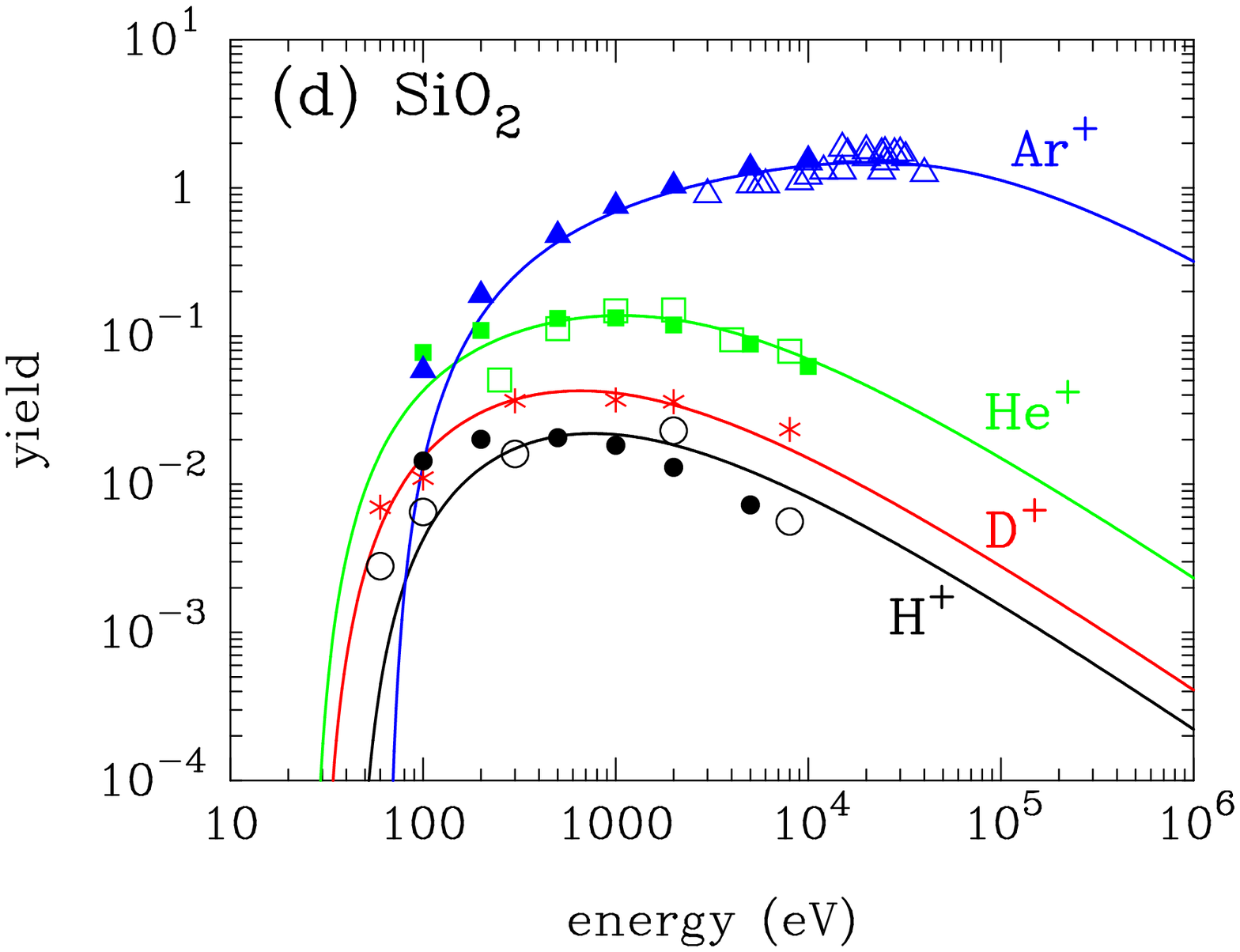}
\plottwo{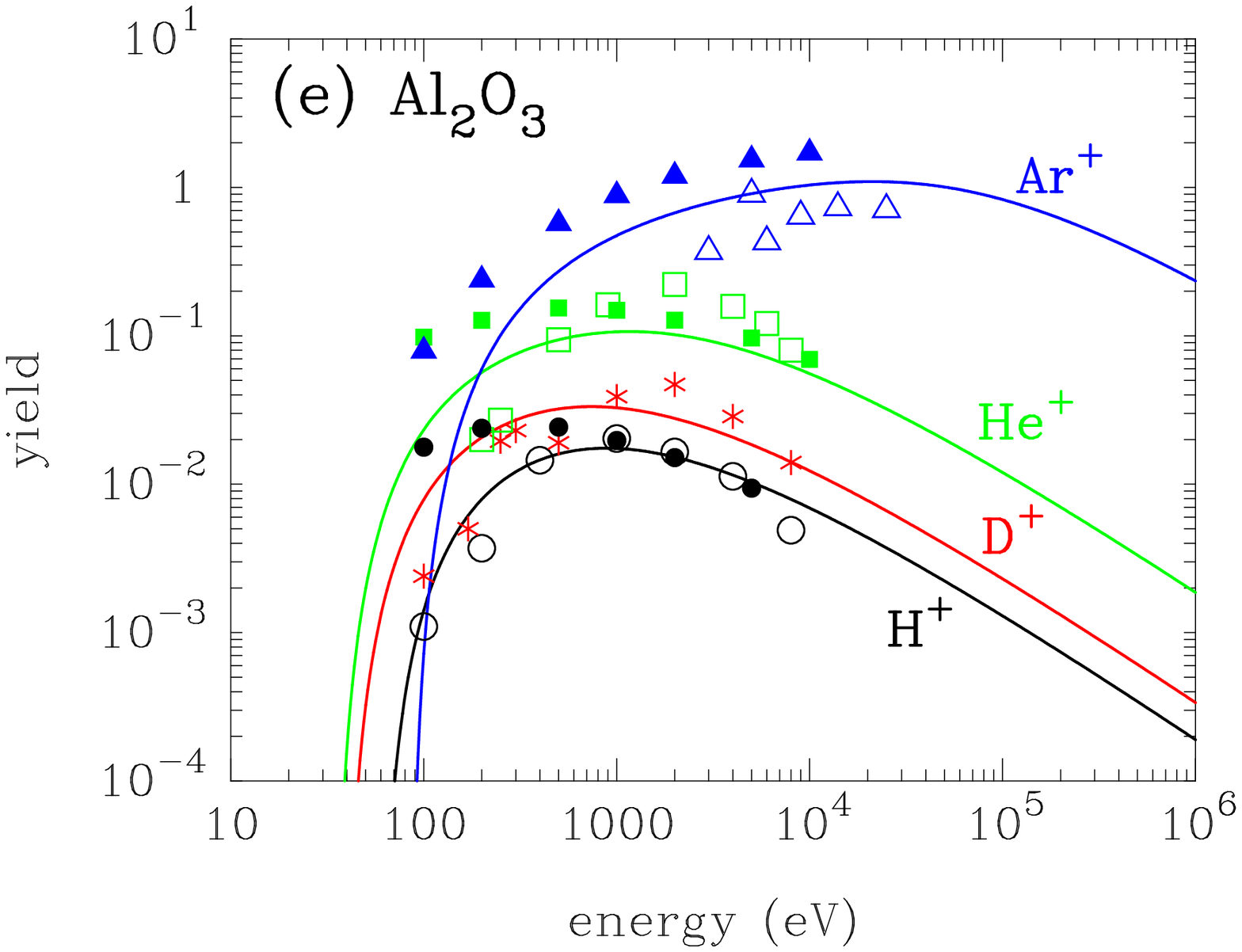}{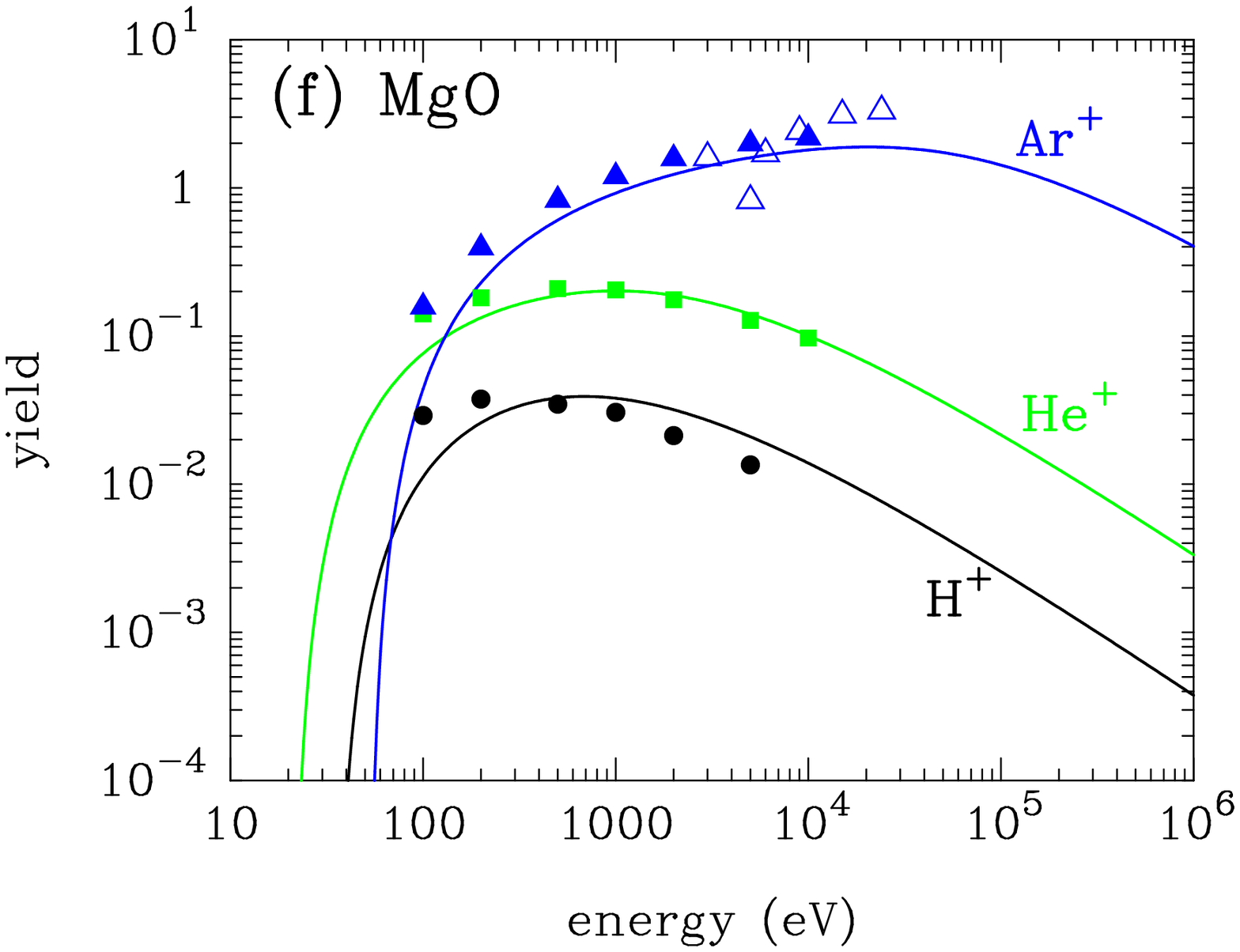}
\plottwo{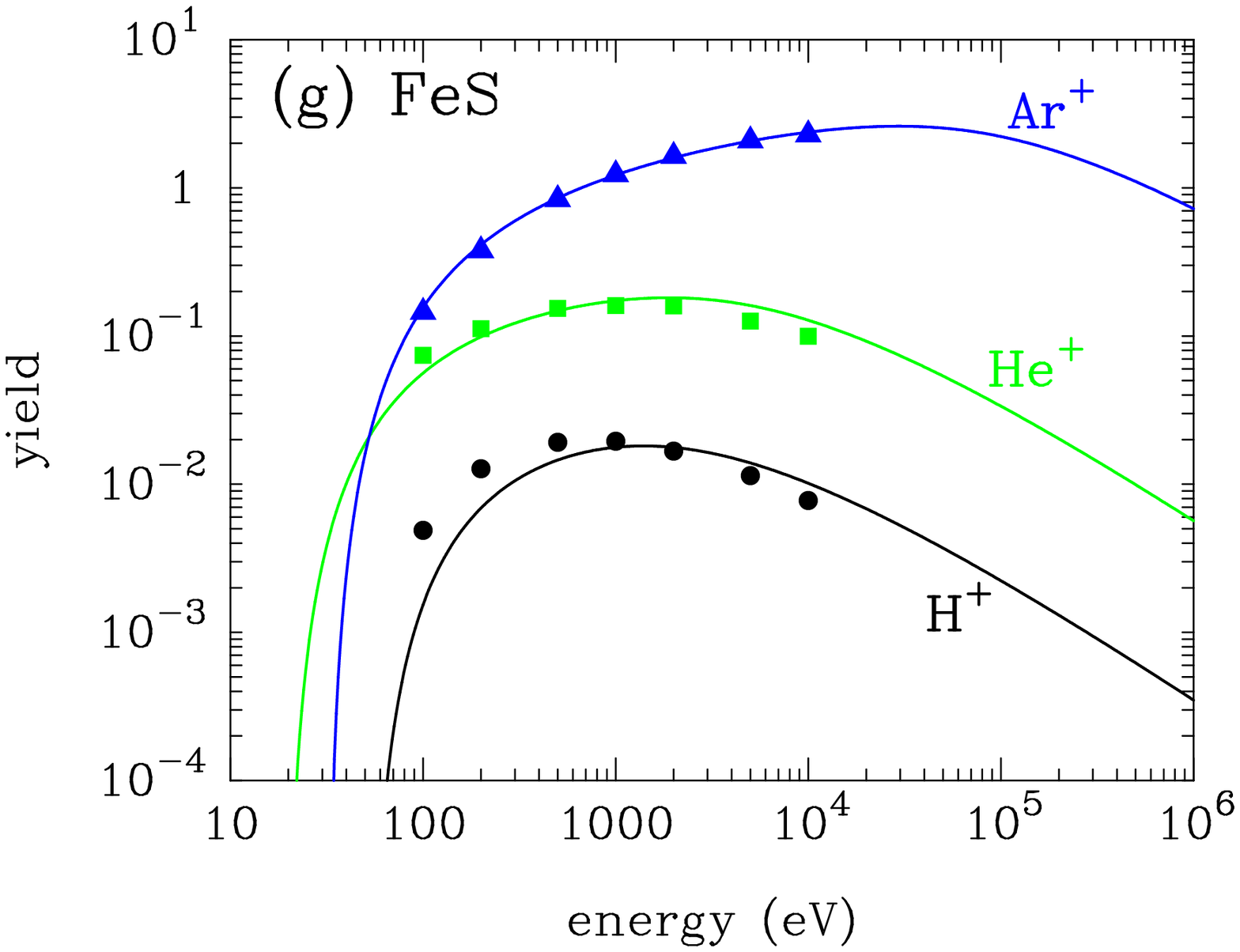}{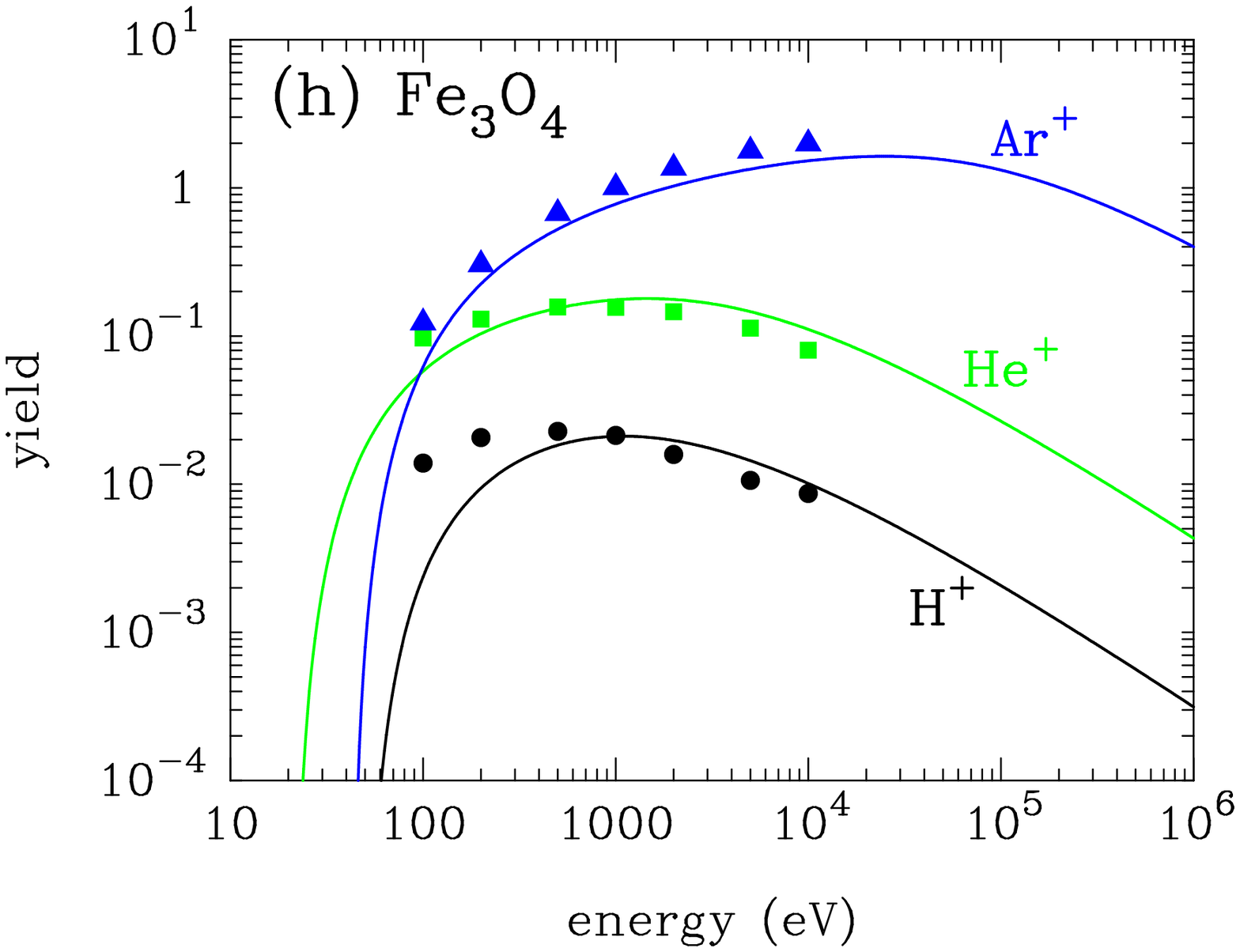}
\caption{\small{ 
 Sputtering yields of each grain species versus incident energy of
 projectiles. 
 (a) C, (b) Si, (c) Fe, (d) SiO$_2$, (e) Al$_2$O$_3$, (f) MgO, (g) FeS, 
 and (h) Fe$_3$O$_4$.
 The incident ion species are H$^+$, D$^+$,  He$^+$, Ne$^+$ and Ar$^+$.
 The experimental data on sputtering yield are represented by open 
 circles (H$^+$), asterisks (D$^+$), open squares (He$^+$), crosses 
 (Ne$^+$) and open triangles (Ar$^+$), and the results of sputtering 
 yield calculated by the EDDY code are denoted by filled circles
 (H$^+$), filled squares (He$^+$) and filled triangles (Ar$^+$).
 The solid curves show the best-fitting theoretical yields calculated by 
 the universal relation.
 In (b) and (c), the dotted curves show the theoretical sputtering
 yields by Ar$^+$ projectile calculated by adopting $\alpha_i$ given by
 Equation (17).
 \textit{See the electronic edition of the Journal for a color version
 of this figure.}\label{fig1}}}
\end{figure}

\clearpage

\begin{figure}
\epsscale{0.7}
\plotone{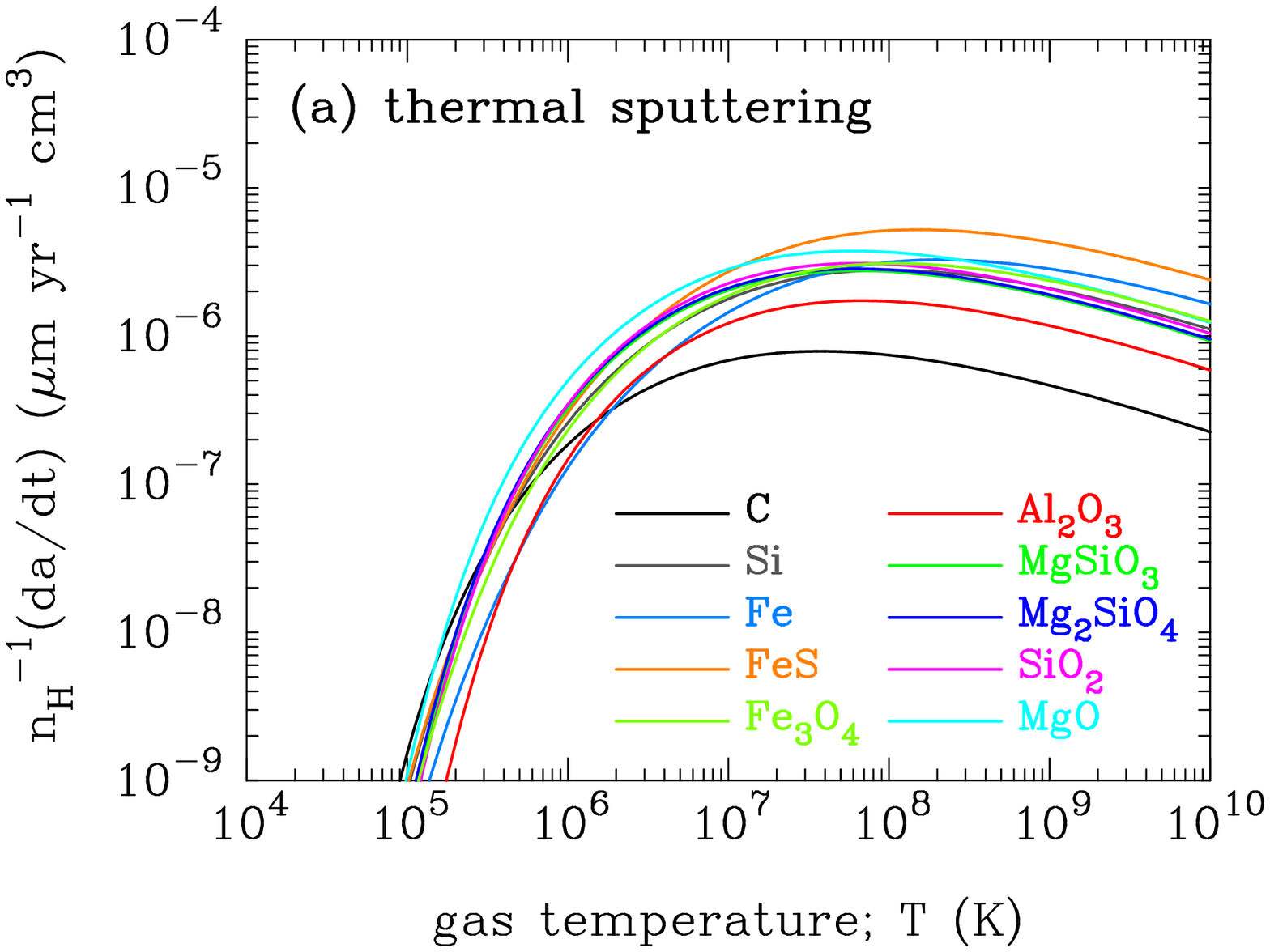}
\plotone{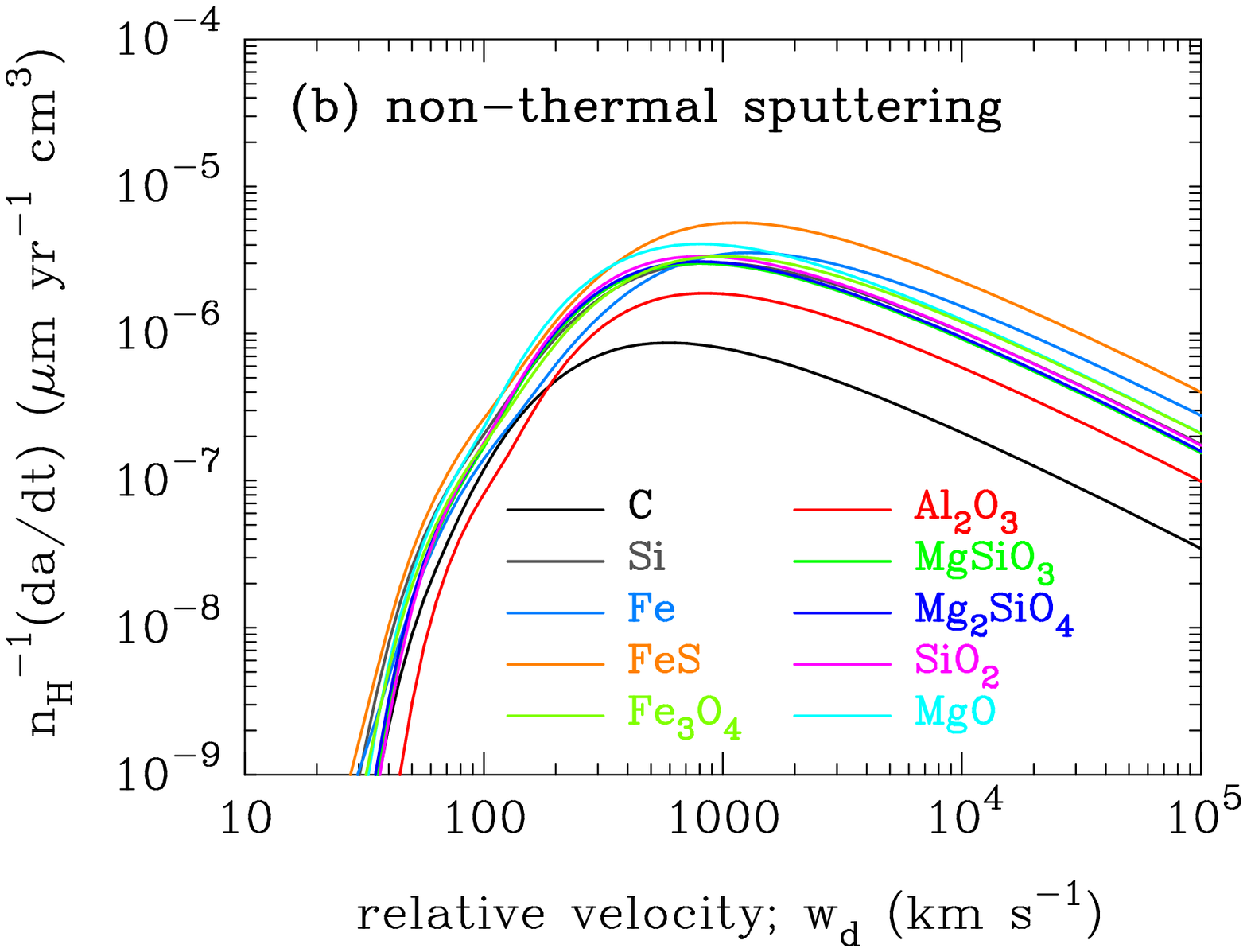}
\caption{
 The erosion rate of each dust species by sputtering in units of $\mu$m 
 yr$^{-1}$ cm$^3$ calculated for the elemental composition of gas with 
 $Z = 10^{-4}$ $Z_\odot$ given in Table 1;
 (a) the thermal sputtering calculated by Equation (22) as a function of 
 gas temperature $T$ and (b) the non-thermal sputtering calculated by 
 Equation (23) as a function of velocity of dust relative to gas 
 $w_{\rm d}$.
 \textit{See the electronic edition of the Journal for a color version 
 of this figure.}\label{fig2}}
\end{figure}

\clearpage

\begin{figure}
\epsscale{0.7}
\plotone{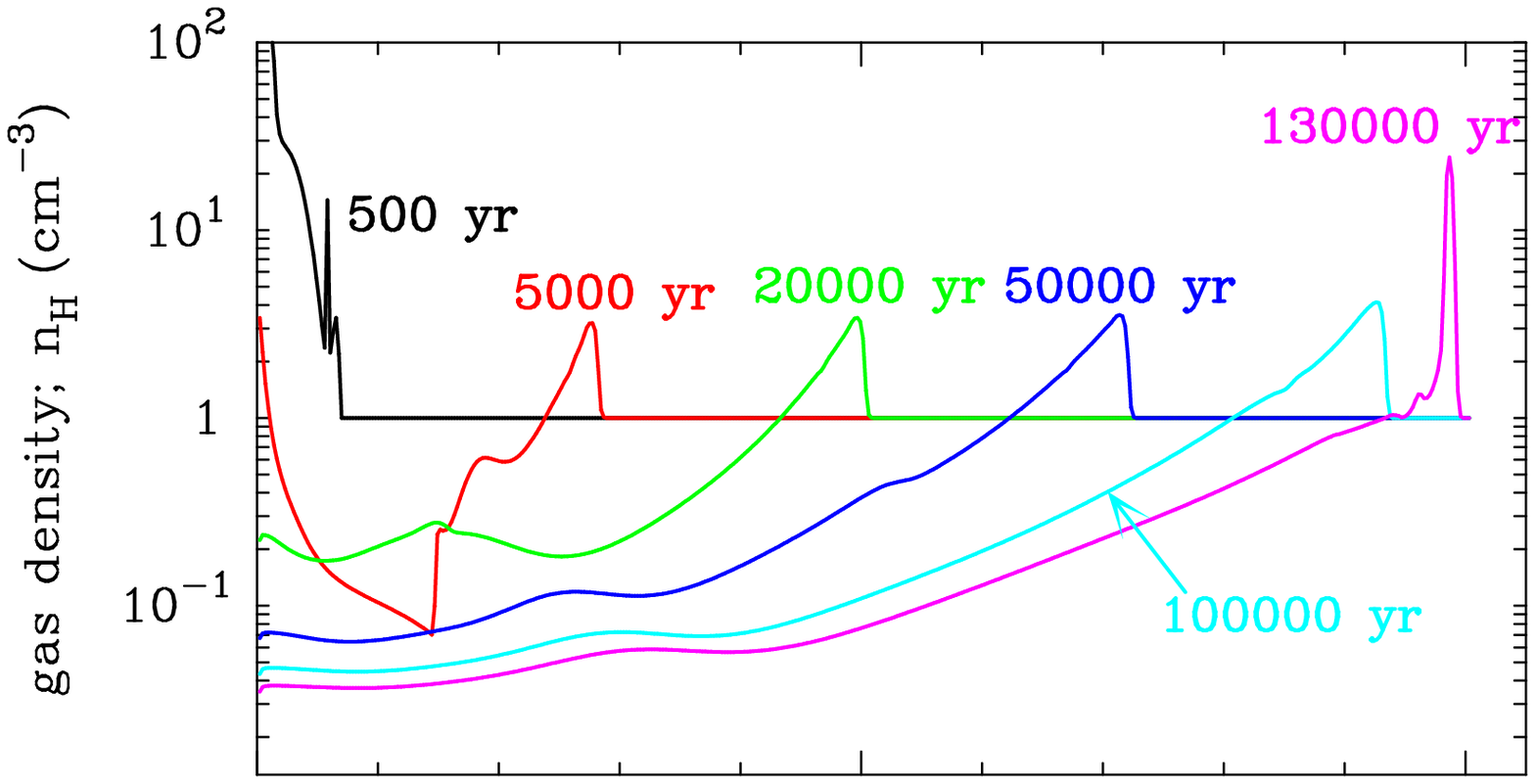}
\plotone{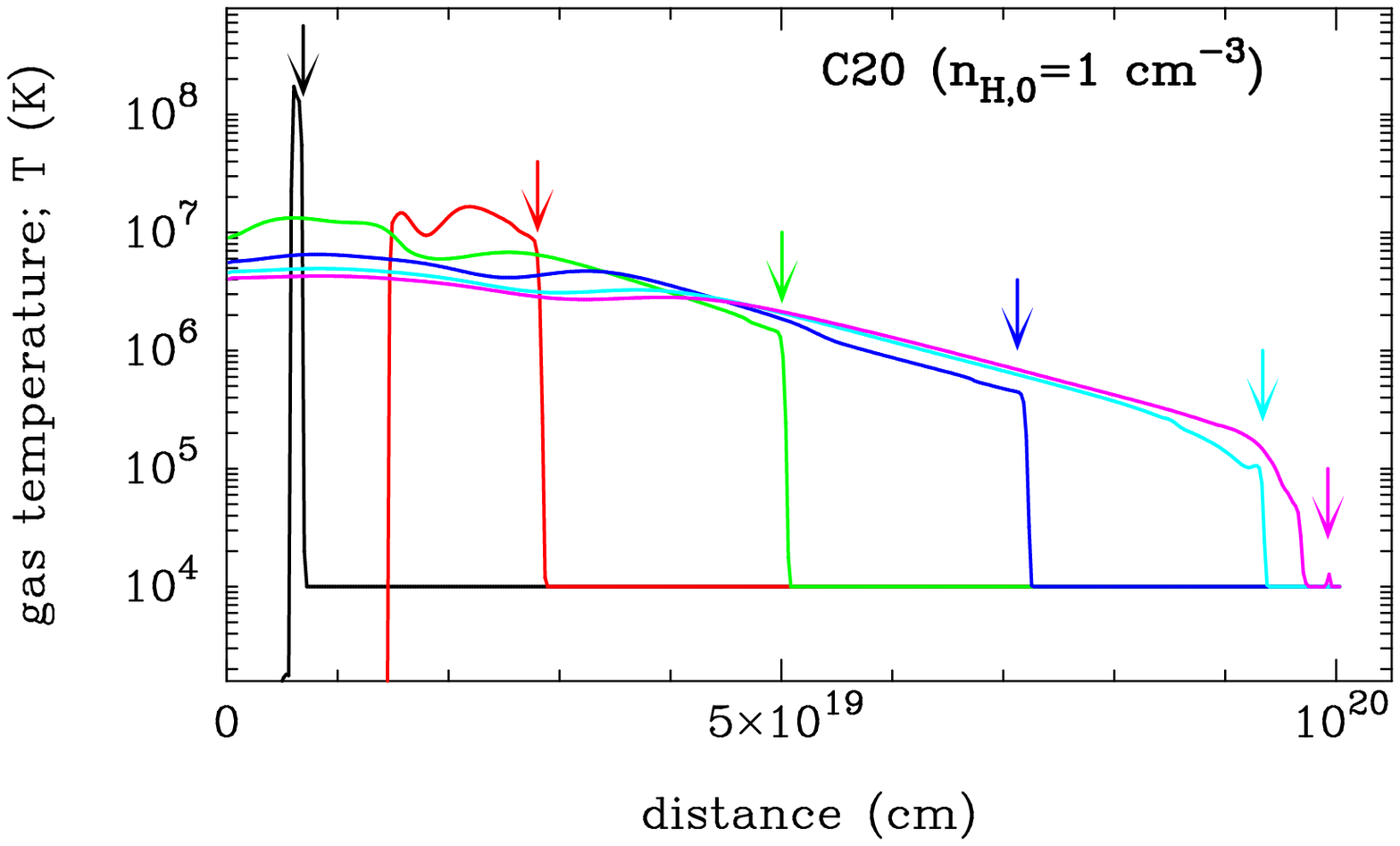}
\caption{ 
 The structures of density (upper panel) and temperature of gas (lower 
 panel) at given times in the interstellar shock that is driven by the
 SN model of C20 with $E_{51}=1$ and $M_{\rm pr} = 20$ $M_\odot$ and is 
 propagating into the ISM with $n_{\rm H, 0} = 1$ cm$^{-3}$.
 The arrow in the lower panel indicates the position of the shock front 
 at a given time.
 \textit{See the electronic edition of the Journal for a color version 
 of this figure.}\label{fig3}}
\end{figure}

\clearpage

\begin{figure}
\epsscale{0.7}
\plotone{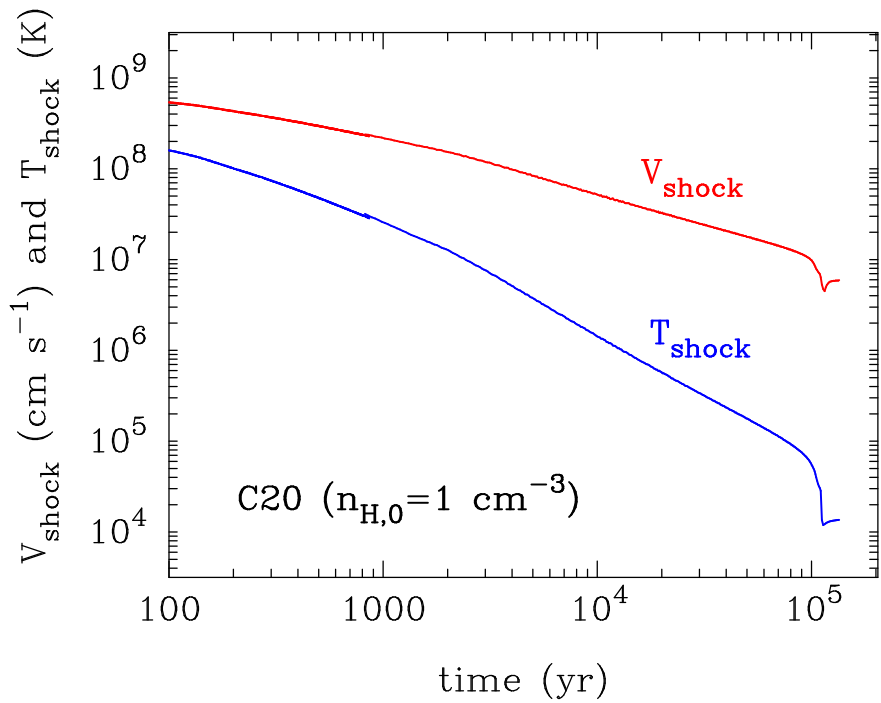}
\caption{
 The time evolution of the shock velocity $V_{\rm shock}$ and the gas 
 temperature $T_{\rm shock}$ at the shock front for model C20 with 
 $n_{\rm H, 0}=1$ cm$^{-3}$.
 \textit{See the electronic edition of the Journal for a color version 
 of this figure.}\label{fig4}}
\end{figure}

\clearpage

\begin{figure}
\epsscale{0.7}
\plotone{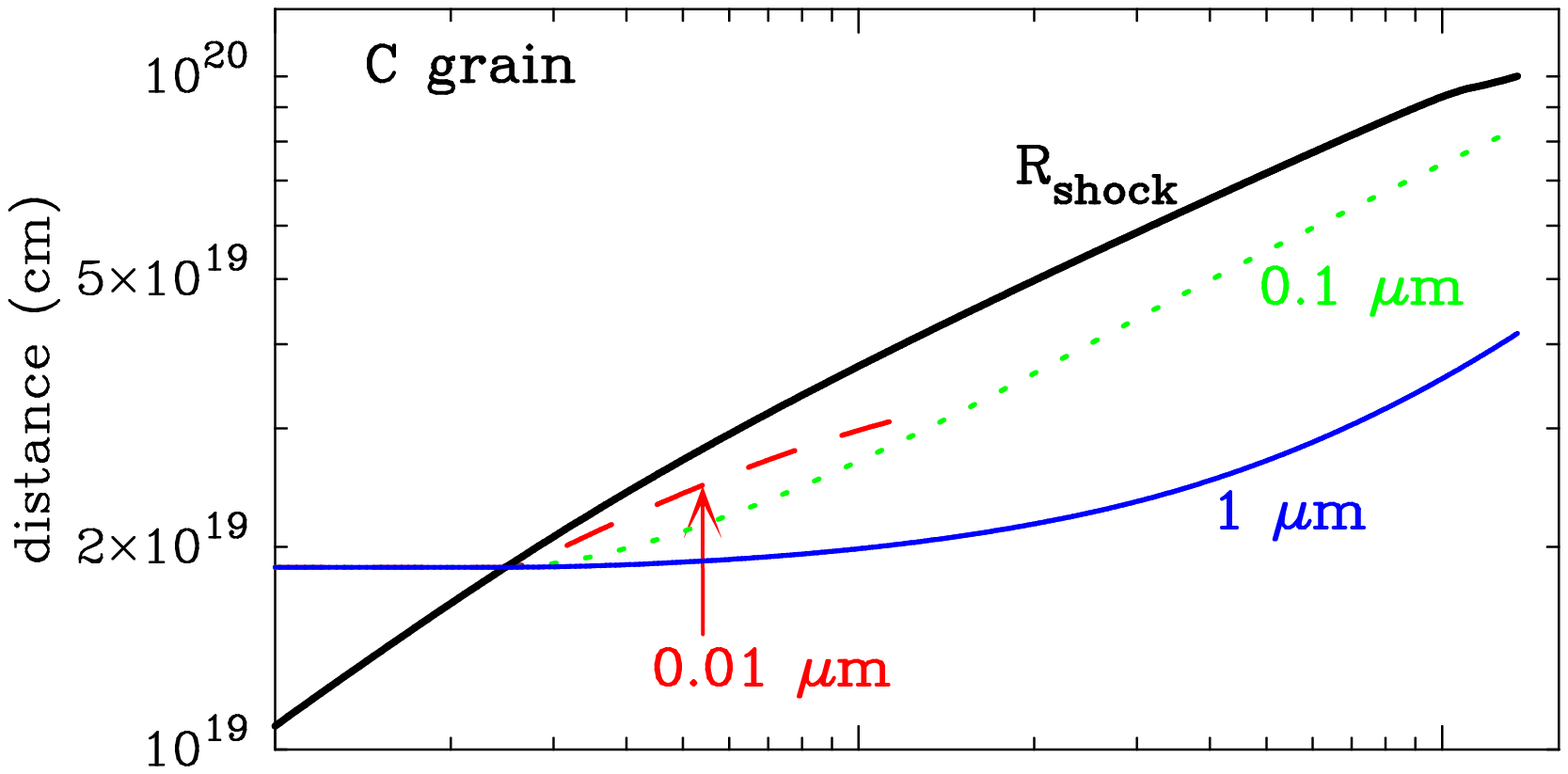}
\plotone{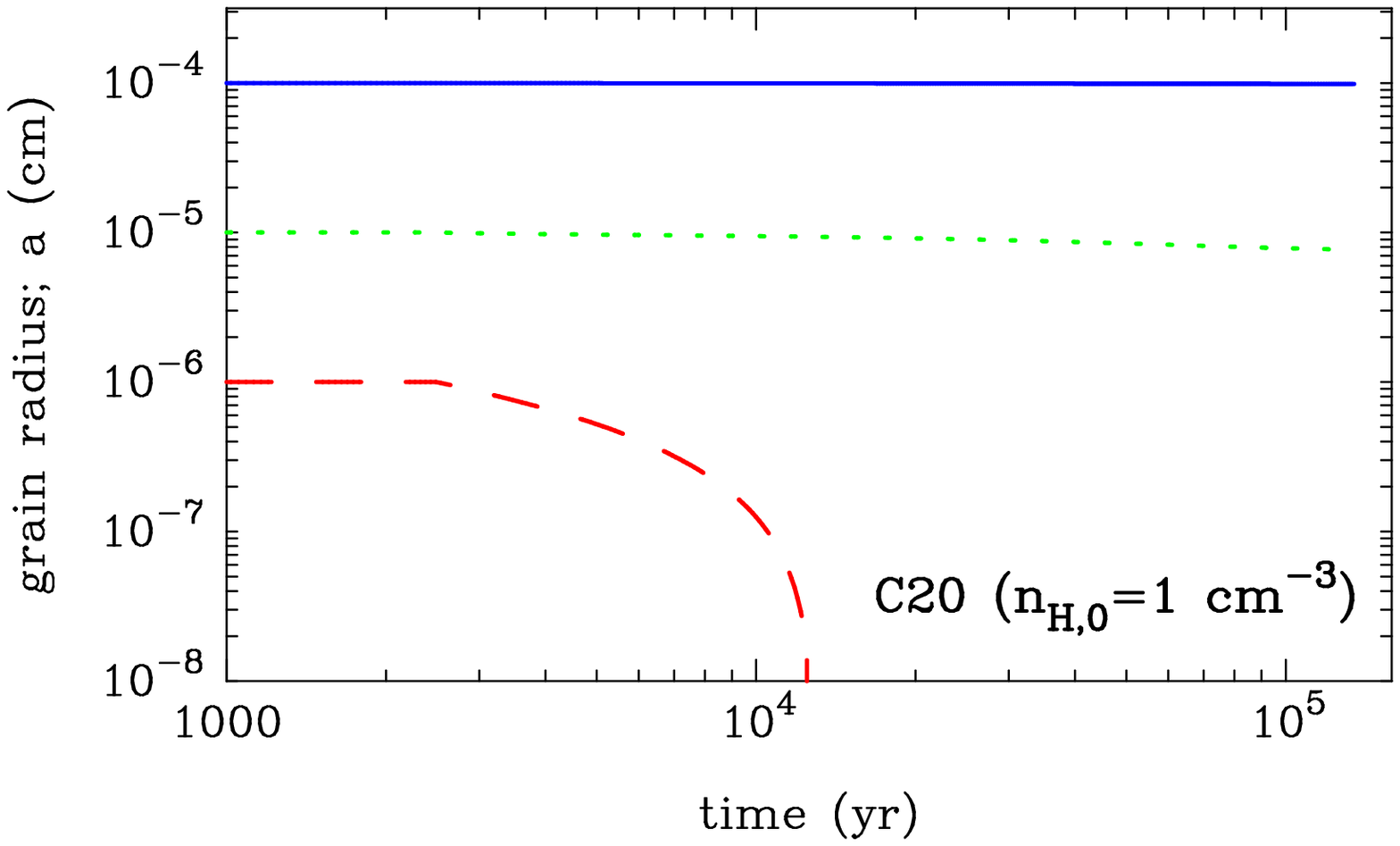}
\caption{ 
 The time evolution of position (upper panel) and size (lower panel) of 
 dust grain in postshock flow for model C20 with $n_{\rm H, 0}=1$ 
 cm$^{-3}$.
 Dust species considered is C grain with the size of 0.01 $\mu$m (dashed), 
 0.1 $\mu$m (dotted) and 1 $\mu$m (solid).
 The thick solid curve in the upper panel denotes the position of the
 shock front $R_{\rm shock}$.
 \textit{See the electronic edition of the Journal for a color version 
 of this figure.}\label{fig5}}
\end{figure}

\clearpage

\begin{figure}
\epsscale{0.7}
\plotone{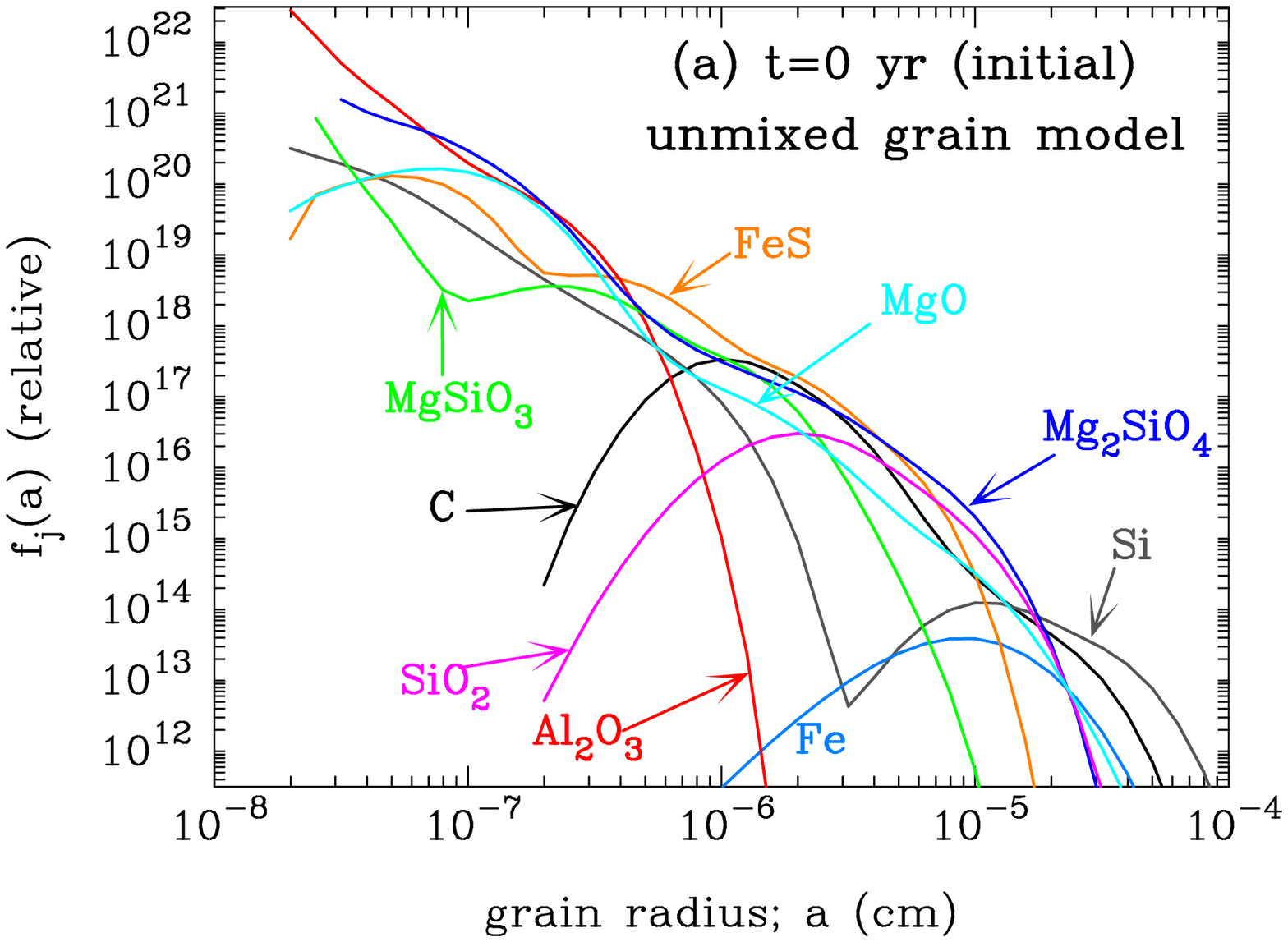}
\plotone{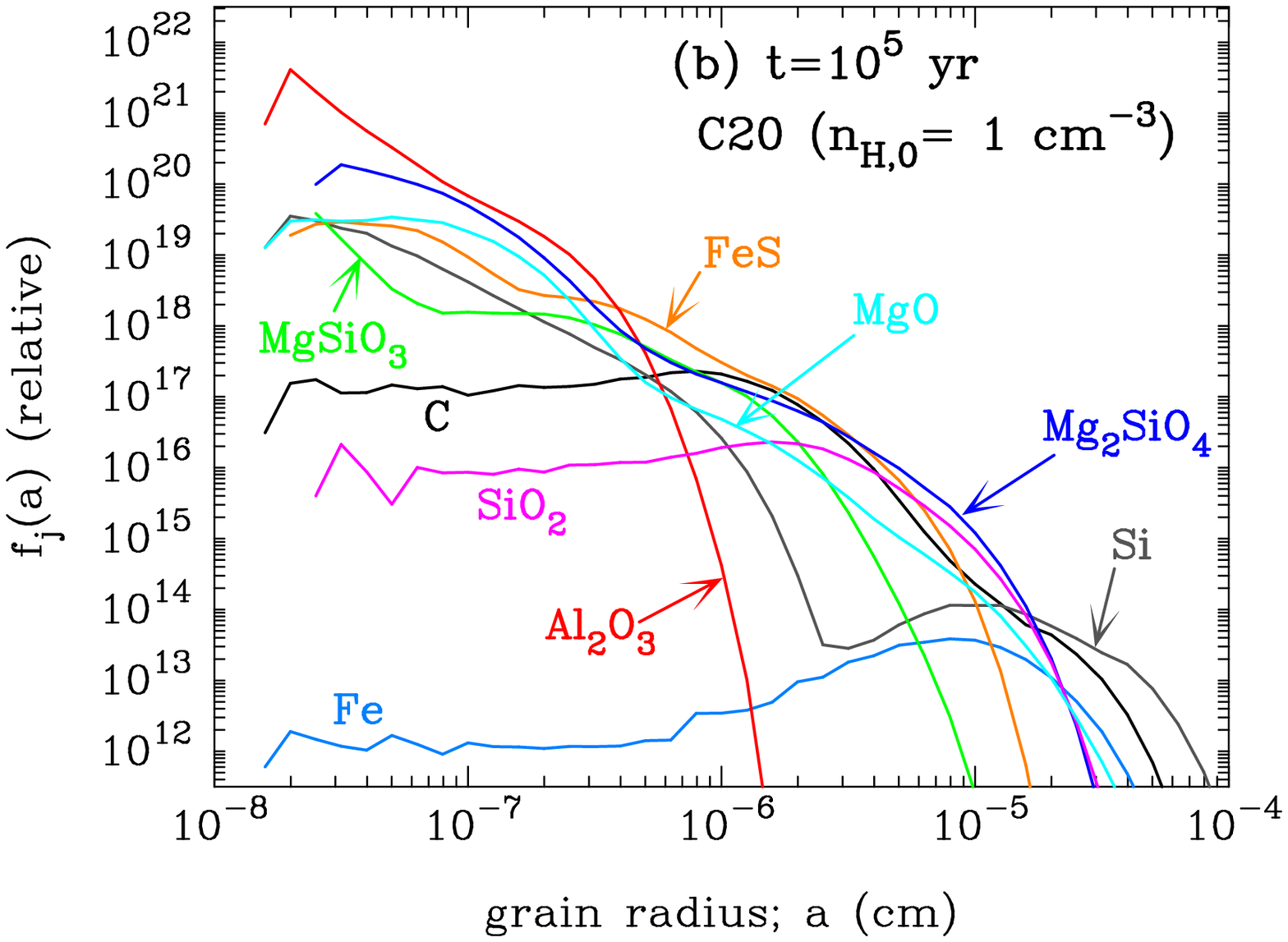}
\caption{ 
 The size distribution of each dust species for the unmixed grain model;
 (a) for the initial size distribution before destruction and 
 (b) for the size distribution obtained by the calculation of dust
 destruction for model C20 with $n_{\rm H, 0}=1$ cm$^{-3}$.
 \textit{See the electronic edition of the Journal for a color version 
 of this figure.}\label{fig6}}
\end{figure}

\clearpage

\begin{figure}
\epsscale{0.7}
\plotone{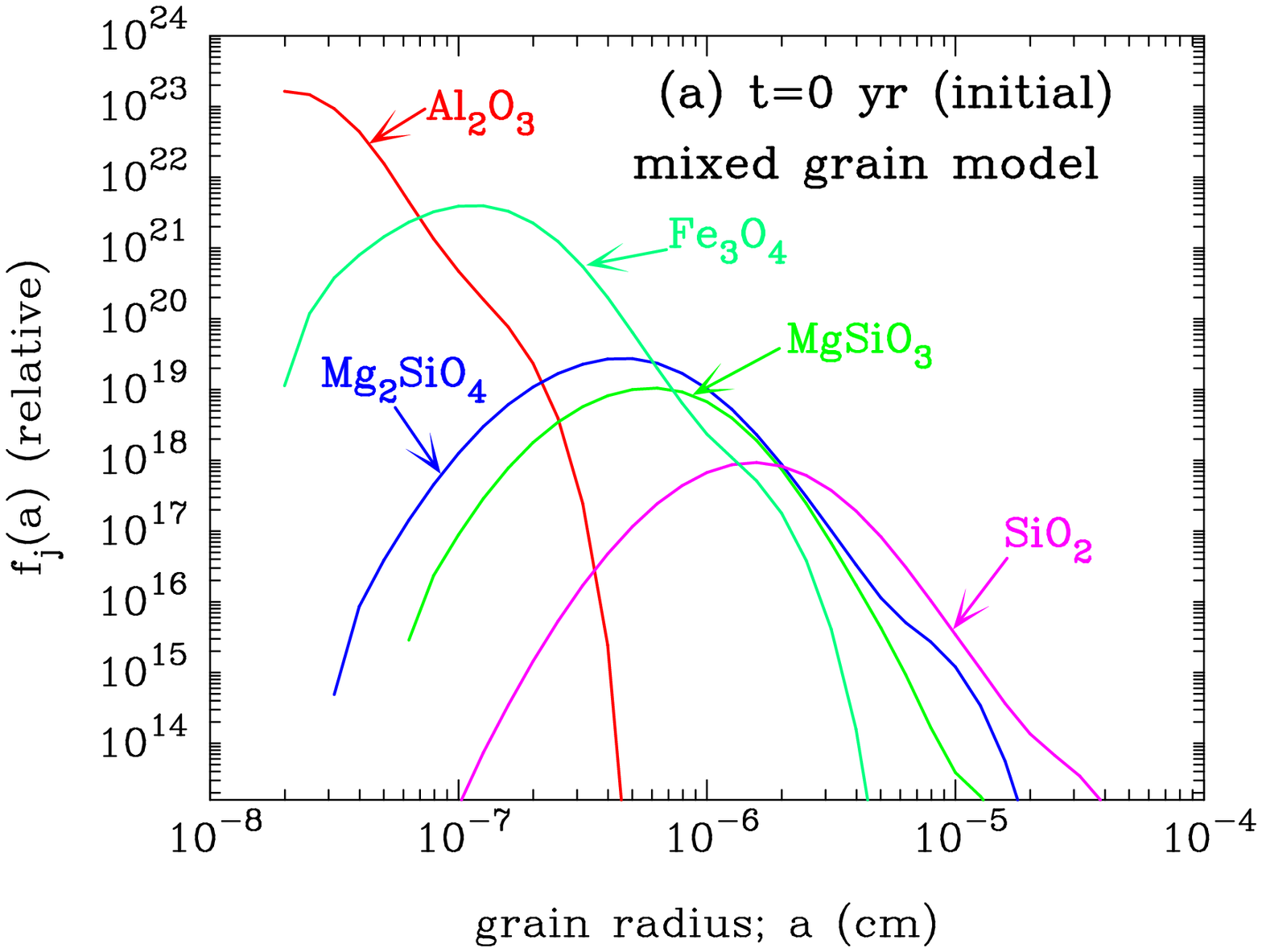}
\plotone{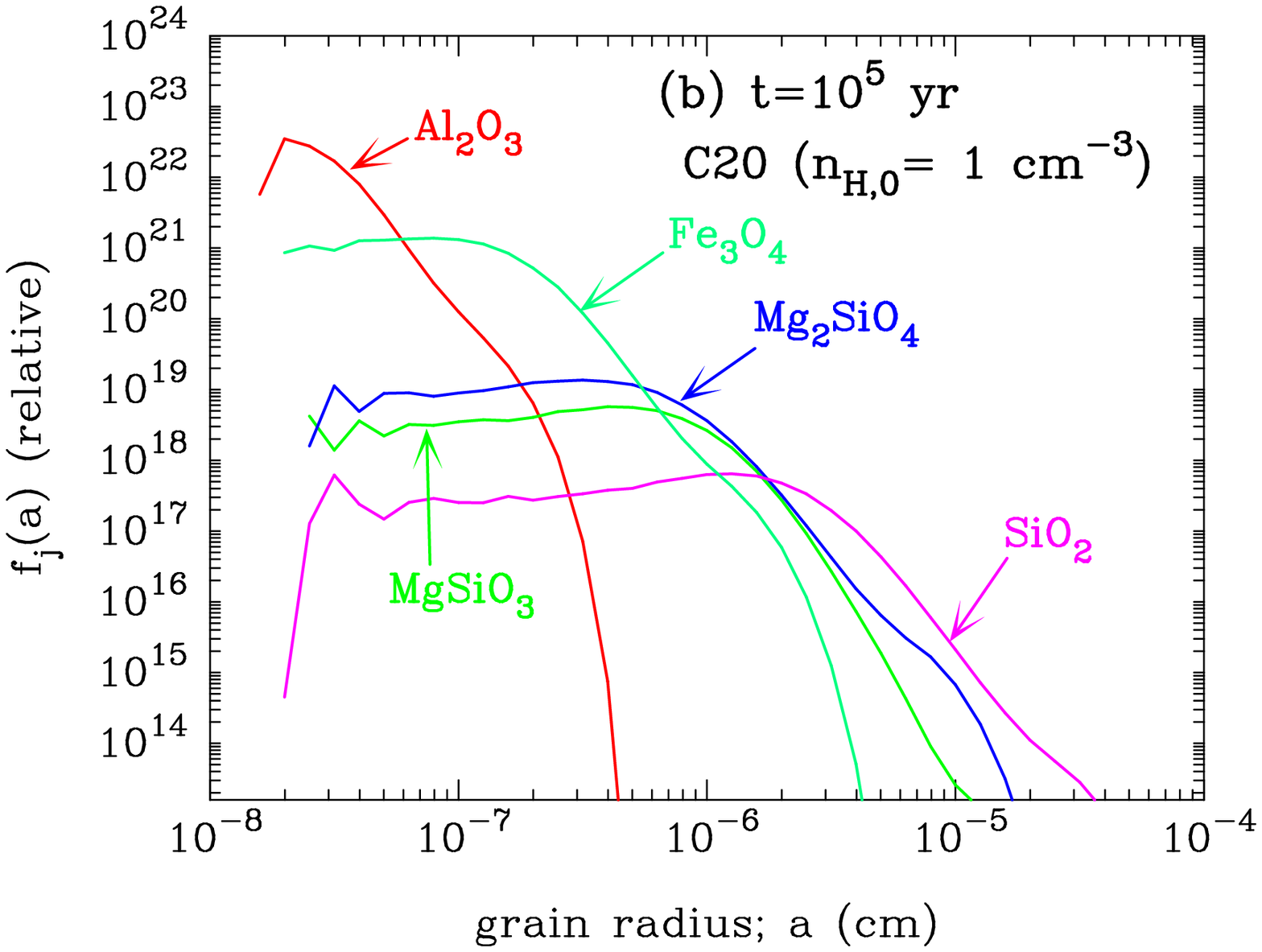}
\caption{ 
 The size distribution of each dust species for the mixed grain model;
 (a) for the initial size distribution before destruction and 
 (b) for the size distribution obtained by the calculation of dust
 destruction for model C20 with $n_{\rm H, 0}=1$ cm$^{-3}$.
 \textit{See the electronic edition of the Journal for a color version 
 of this figure.}\label{fig7}}
\end{figure}

\clearpage

\begin{figure}
\epsscale{0.7}
\plotone{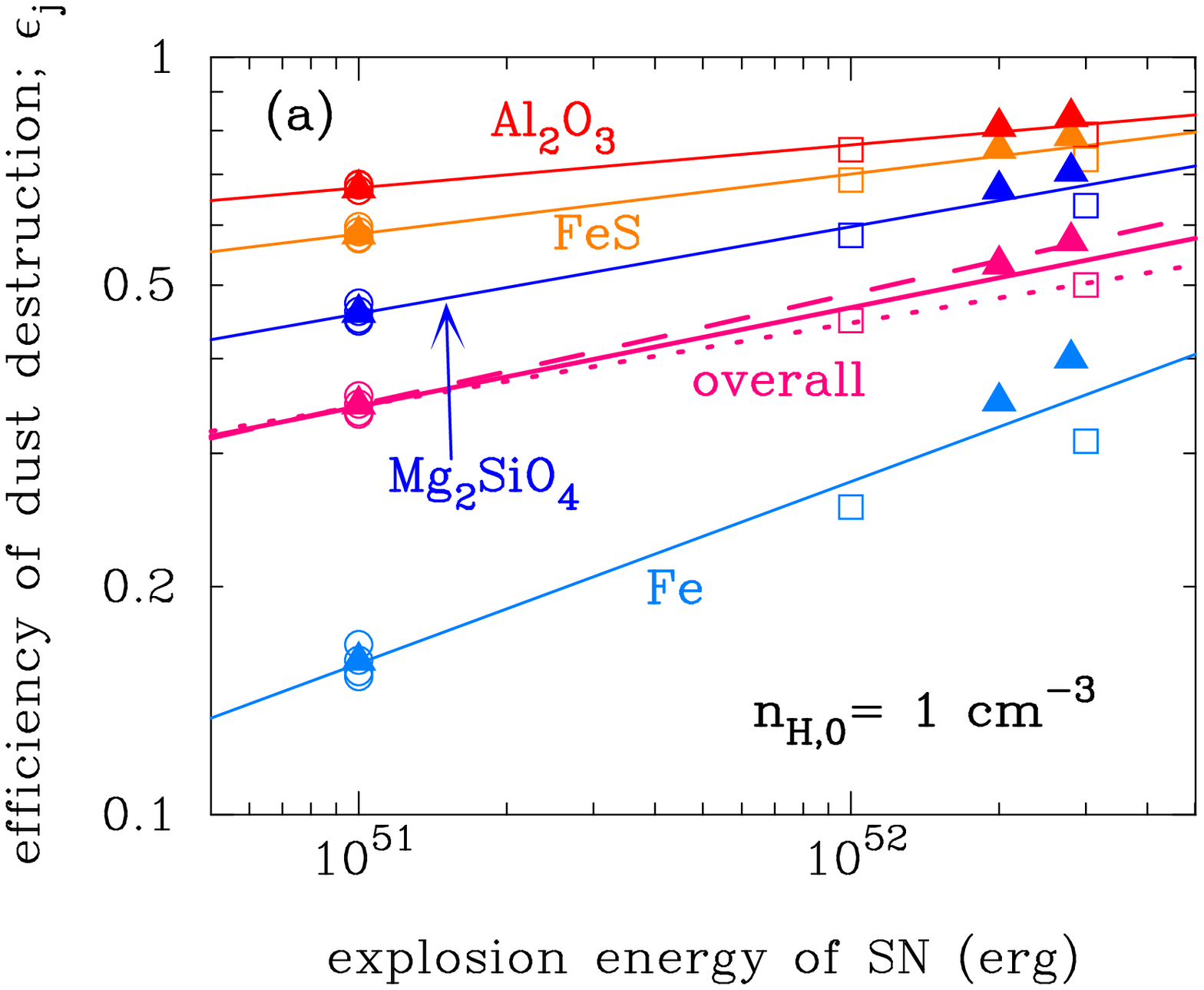}
\plotone{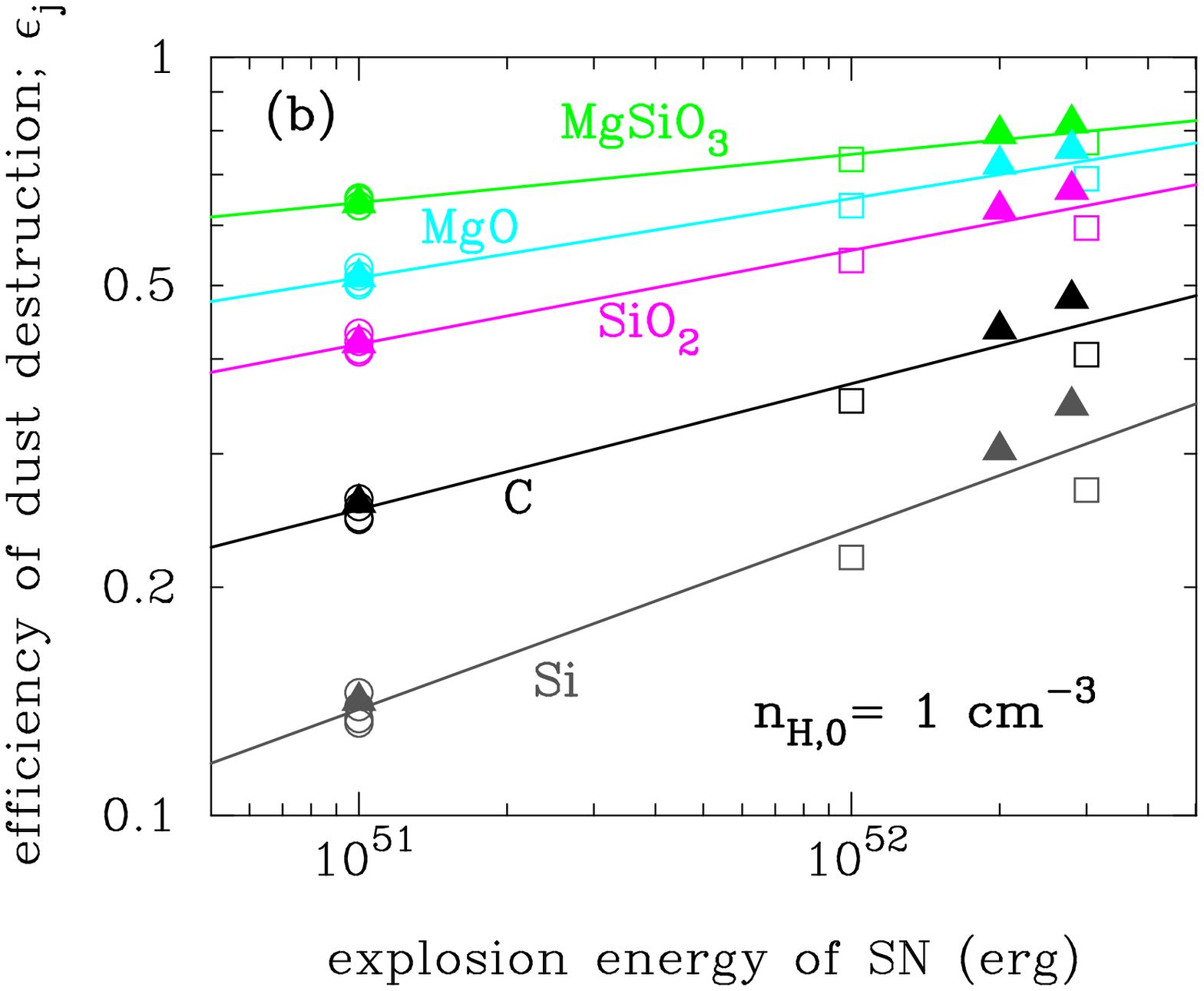}
\caption{\small{
 The efficiency of destruction of each grain species in the unmixed
 grain model calculated for $n_{\rm H, 0}=1$ cm$^{-3}$ as a 
 function of SN explosion energy; (a) for Al$_2$O$_3$, FeS,
 Mg$_2$SiO$_4$ and Fe grains and (b) for MgSiO$_3$, MgO, SiO$_2$, C and 
 Si grains. The SN models used for the calcularions are distinguished by open 
 circles (CCSNe), open squares (HNe) and filled triangles (PISNe).  
 Also, in (a), the overall efficiency of dust destruction is plotted.
 The linear solid lines for each grains species are the results 
 calculated by the power-law formula given by Equation (A1) for all SNe. 
 The dotted and dashed lines for the overall efficiencies of dust 
 destruction are the results of calculations for SNe II and for PISNe, 
 respectively.
 \textit{See the electronic edition of the Journal for a color version 
 of this figure.}\label{fig8}}}
\end{figure}

\clearpage

\begin{figure}
\epsscale{0.7}
\plotone{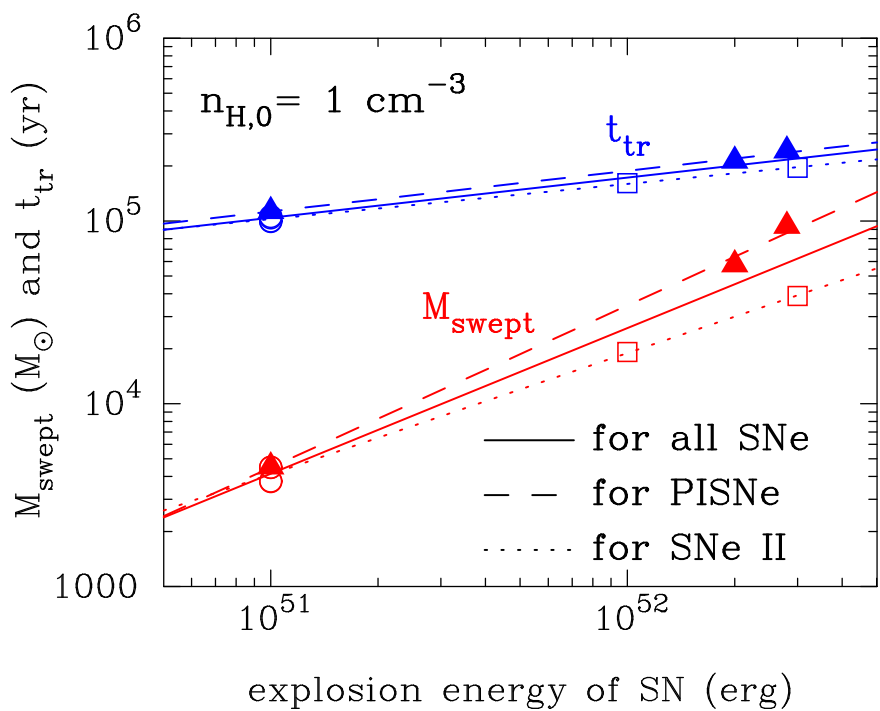}
\caption{
 The mass of gas swept up by shock $M_{\rm swept}$ and the truncation 
 time $t_{\rm tr}$ versus the explosion energy of SN for 
 $n_{\rm H, 0}=1$ cm$^{-3}$.
 The results for CCSNe, HNe and PISNe are represented by open circles, 
 open squares and filled triangles, respectively.
 The linear lines are the power-law formula approximated by
 Equation (A2) for all SNe (solid), SNe II (dotted) and PISNe (dashed).
 \textit{See the electronic edition of the Journal for a color version 
 of this figure}.\label{fig9}}
\end{figure}

\clearpage

\begin{figure}
\epsscale{0.7}
\plotone{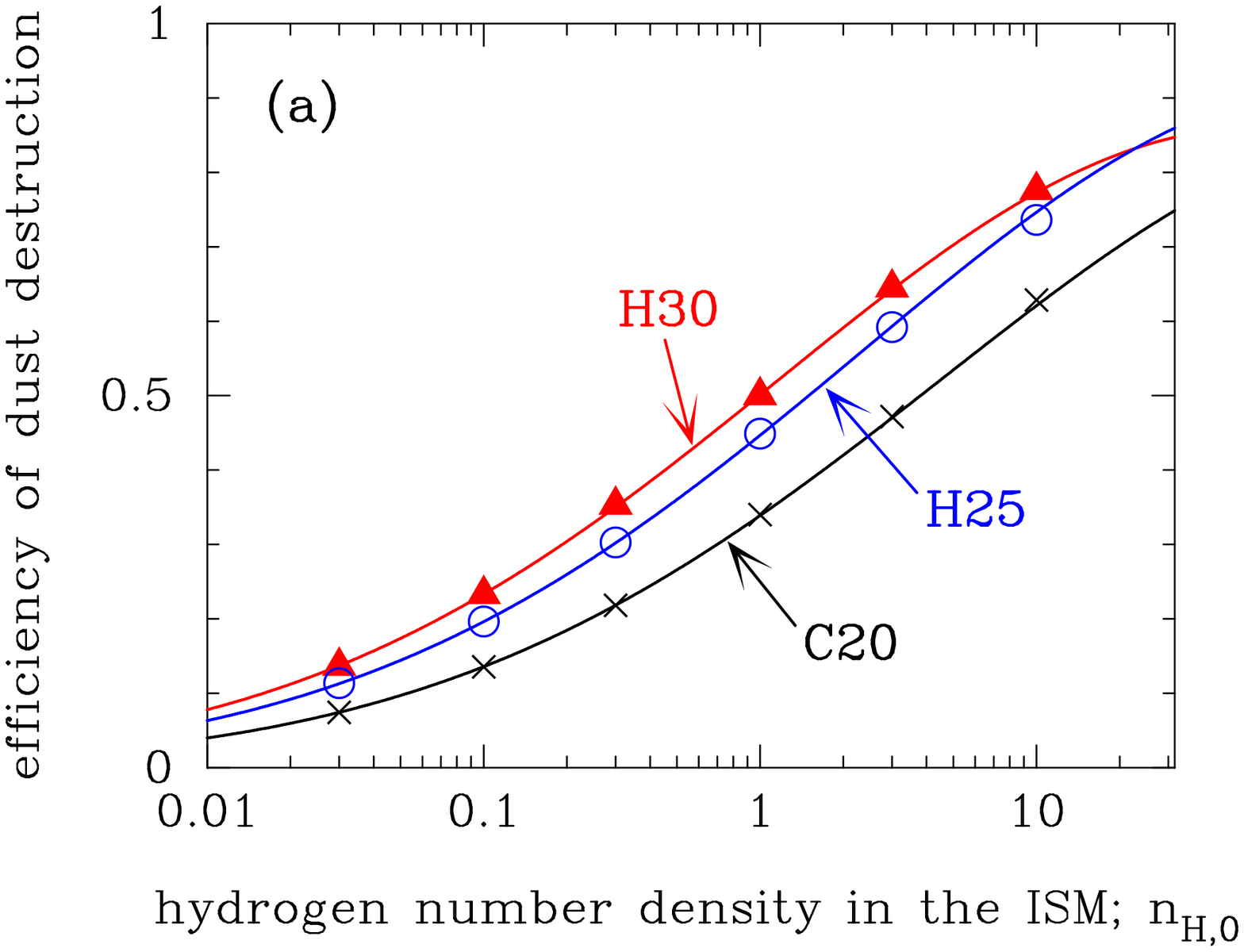}
\plotone{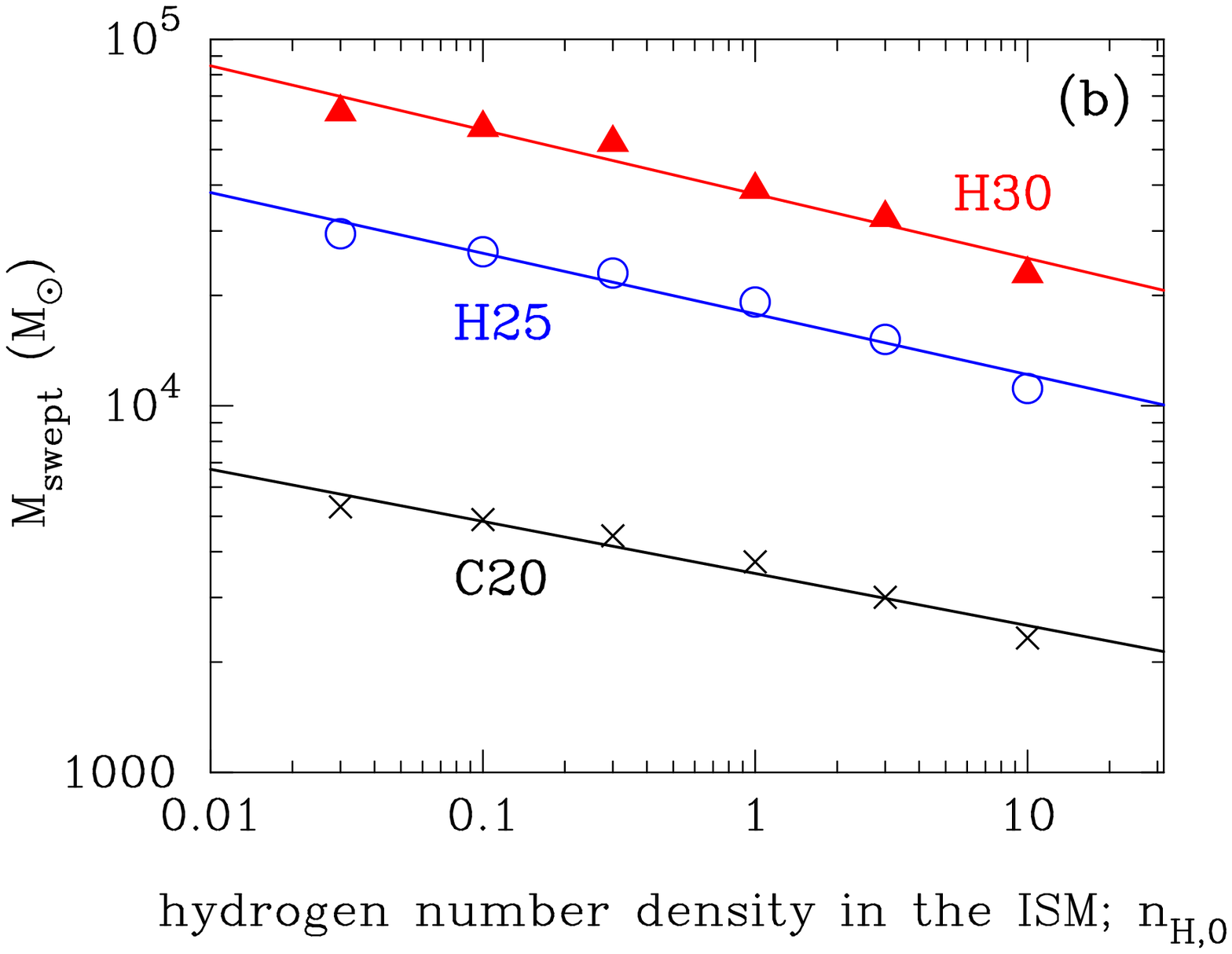}
\caption{
 (a) The overall efficiency of dust destruction versus $n_{\rm H, 0}$ 
 for C20 (crosses), H25 (open circles) and H30 models (filled 
 triangles).
 The solid curves are calculated by the approximation formula by Equation
 (A3).
 (b) The mass of gas swept up by shock for C20 (crosses), H25 (open
 circles)  and H30 models (filled triangles) as a function of 
 $n_{\rm H, 0}$.
 The linear lines are the power-law approximation formula of 
 $M_{\rm swept} \propto n_{\rm H, 0}^g $ with $g = -0.142 E_{51}^{0.063}$.
 \textit{See the electronic edition of the Journal for a color version 
 of this figure.}\label{fig10}}
\end{figure}

\clearpage

\begin{figure}
\epsscale{0.7}
\plotone{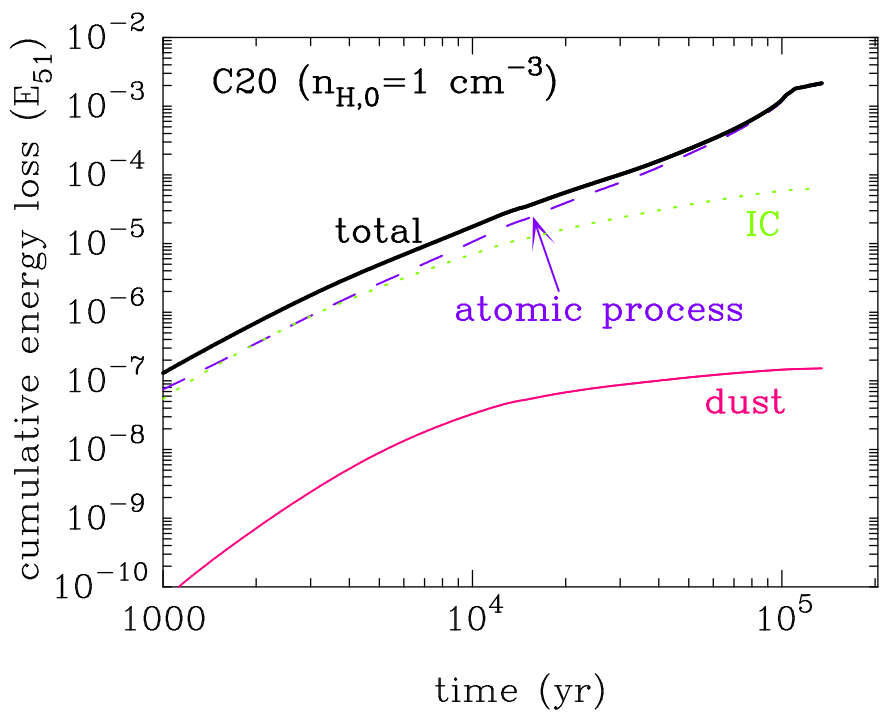}
\caption{
 The time evolution of cumulative energy lost by the atomic process
 (dashed), the inverse Compton cooling (dotted) and the thermal emission 
 from all dust grains (solid) in units of $E_{51}$ for the SN model C20,
 $n_{{\rm H}, 0} = 1$ and $Z = 10^{-4}$ $Z_\odot$ at the redshift of $z=20$. 
 The total energy loss is indicated by the thick solid curve.
 \textit{See the electronic edition of the Journal for a color version 
 of this figure.}\label{fig10}}
\end{figure}

\end{document}